\documentclass[%preprint, linenumbers,%groupedaddress,%unsortedaddress,%runinaddress,%frontmatterverbose,%preprint,%preprintnumbers,%nobibnotes,%bibnotes,%pra,%prb,%rmp,%prstab,%prstper,
reprint,superscriptaddress,floatfix,amsmath,amssymb,aps,prc,nofootinbib]{revtex4-2}
\usepackage{graphicx}% Include figure files
\usepackage{dcolumn}% Align table columns on decimal point
\usepackage{bm}% bold math
\usepackage{fourier,physics,slashed,tensor}
\usepackage{diagbox}
\usepackage{subfigure}
\usepackage{soul}
\usepackage{tabularx}
\usepackage{float}
\usepackage{epstopdf}
\usepackage{epsfig}

\newcommand{\ee}{\end{eqnarray}}

\newcommand{\eqcomma}{\phantom{AA},\phantom{AA}}
 \newcommand{\eqn}[1]{Eq.\,(\ref{#1})}
\usepackage{xcolor}
\usepackage{comment}
\usepackage[normalem]{ulem}  % \sout{old text} for strikeout
\renewcommand\sout{\bgroup \color{red} \ULdepth=-.5ex \ULset}
\begin{document}

\title{\textbf{
$f_2(1270)\to\pi+\pi$ as a Probe of Spin and Vorticity in Heavy-Ion Collisions}}

\author{In Woo Park}
\email{inwoopark95@gmail.com}
\affiliation{Department of Physics and Institute of Physics and Applied Physics, Yonsei University, Seoul, Korea}

\author{Beomkyu Kim}
\email{kimbyumkyu@skku.edu}
\affiliation{Department of Physics, Sungkyunkwan University, Suwon, Korea}

\author{Giorgio Torrieri}
\email{torrieri@unicamp.br}
\affiliation{Universidade Estadual de Campinas, Instituto de Fisica “Gleb Wataghin” Rua Sergio Buarque de Holanda 777, Campinas, Sao Paulo, Brazil}

\author{Kayman J. Gonçalves}
\email{kaymanjhosef10@gmail.com}
\affiliation{Universidade Estadual de Campinas, Instituto de Fisica “Gleb Wataghin” Rua Sergio Buarque de Holanda 777, Campinas, Sao Paulo, Brazil}

\author{Sanghoon Lim}
\email{shlim@pusan.ac.kr}
\affiliation{Department of Physics, Pusan National University, Pusan, Korea}

\author{Su Houng Lee}
\email{suhoung@yonsei.ac.kr}
\affiliation{Department of Physics and Institute of Physics and Applied Physics, Yonsei University, Seoul, Korea}

\date{\today}% It is always \today, today,

\begin{abstract}
The correlation between vorticity and spin alignment in heavy-ion collisions can be probed through polarization measurements of hadrons, whose total spin originates from both constituent-quark spins and orbital angular momentum in the quark-model framework. 
To motivate such experimental studies, we calculate the general angular distribution of produced pion in $f_2(1270)\to\pi+\pi$ using interaction Lagrangian and helicity formalism and check that both methods yield the same result. The distribution is given as a function of angle between pion and initial quantization axis of $f_2$ and the spin density matrix element of $f_2$. Its diagonal entries and $\rho_{20}$ component were computed assuming local thermal equilibrium and blast wave model for different centrality classes, hence given as a function of azimuthal angle with respect to the impact parameter.
\end{abstract}

\maketitle
\section{\label{sec:intro}Introduction}

In non-central heavy ion collisions, a huge global orbital angular momentum (OAM) of the order of $10^5-10^7\hbar$ is generated. The generated OAM is partially transferred to the spin polarization of quarks and anti-quarks through spin-orbit coupling, resulting in the global polarization of the hadrons along the OAM. The global spin polarization of $\Lambda$ hyperon was measured through a self analyzing weak decay $\Lambda\to p+\pi$ \cite{STAR:2017ckg}. On the other hand, the measurement of vector meson spin alignment was done through its strong \textit{p}-wave decay into two pseudoscalar mesons such as $\phi\to K^+K^-$, $K^*\to K\pi$ \cite{Xia:2020tyd,STAR:2022fan,ALICE:2019aid} and $D^*\to D\pi$ \cite{ALICE:2025cdf}.

In a thermal environment the polarization is given by the trace over the thermalized density matrix
\begin{equation}
\label{density}
P_h =\mathrm{Tr}\left(\hat{S}_{z}.\hat{\rho}\right) \eqcomma \hat{\rho}= \frac{1}{Z}\exp\left[ \frac{\vec{\hat{S}}\vdot\vec{\Omega}}{T}\right]\eqcomma Z = \Tr\left(\hat{\rho}\right)
\end{equation}
where $\hat{\vec{S}}$ is the spin vector in the right representation (spinor, vector, tensor etc), $\vec{\Omega}$ is the vorticity, and we define the $z$ direction as the one we want to measure (usually experimentalists measure the global polarization perpendicular to the impact parameter, local polarization in the beam direction, jet polarization in the jet plane).

Note that \cite{wangspin} weak decays allow us to measure polarization $P_h$, but strong decays are parity-conserving, therefore they only allow us to measure spin alignment, $\sim P_{h}^{2}$. Nevertheless, if the system is in thermal equilibrium, \eqn{density} should allow the calculation of the spin alignment of all particles just from the two parameters of $\Omega$ and $T$.

However, there are very good reasons to think that equilibrium between spin and vorticity is imperfect \cite{ourpaper}. At most, the spin density $s_{\alpha \beta}$ and the vorticity $\Omega_{\alpha \beta}$ combine in a relativistic Fert-Valet type equation governed by a time-scale $\tau_\Omega$, related to the vortical susceptibility $\chi$ by a fluctuation-dissipation type relation \cite{montediss,montediss2,montediss3} given as
\begin{equation}
\label{fertvalet}
\tau_\Omega u_\alpha \partial^\alpha s_{\mu \nu}+s_{\mu \nu}=\chi \Omega_{\mu \nu},
\end{equation}
in which case there are indications \cite{kaminski,montediss3} that $\tau_\Omega$ is of the order of the lifetime of the fireball.

In this case, the formula \eqn{density} cannot be the full story, since spin and angular momentum carried by vorticity somehow combine to produce a hadron density matrix. For a quark model coalescence, conservation of angular momentum imposes a formula \cite{kayman1} dependent on one unknown ``decoherence`` parameter $P_L(\omega)$, the probability of a vortex with vorticity $\omega$ to give the hadron $L$ of angular momentum, and Clebsch-Gordan coefficients and $6-j$ symbols. In the quark coalescence model, conservation of angular momentum leads to an expression \cite{kayman1} that depends on a single unknown decoherence parameter, $P_L(\omega)$. This quantity represents the probability that a vortex with vorticity $\omega$ produces a hadron with total angular momentum $L$. The resulting formula involves appropriate Clebsch–Gordan coefficients and Wigner $6-j$ symbols. For a meson this will be 
\begin{eqnarray}\label{coalmeson}
\left( \hat{\rho}^M \right)^{L}_{m',m'''}&=&\sum_{m'' , m}\sum_{m'_1,m'_2,m'_L}\sum_{m_1,m_2,m_L} P_L(\omega)   d^{j}_{m' m'' }(\nu)\nonumber\\
&&C^{j, m''}_{S_1+S_2, m'_1+m'_2, L, m'_L}C^{S_1+S_2,m'_1+m'_2}_{S_1,m'_1,S_2,m'_2}\nonumber\\
&&C^{j, m}_{S_1+S_2, m_1+m_2, L, m_L}C^{S_1+S_2,m_1+m_2}_{S_1,m_1,S_2,m_2}\nonumber\\
&&\rho^{1}_{m_1,m'_1}(\Omega)\rho^{2}_{m_2,m'_2} (\Omega)\left[d^{j}_{m m'''}(\nu)\right]^{-1},
\end{eqnarray}
here $\nu$ is the coalescence angle, and the {\em quark} density matrices are given by \eqn{density}. $d^j(\nu)$ is Wigner d-matrix of spin $j$. For baryons, the formula is more complicated but of the same type, see \cite{ryb} for details. The diagonal elements of the density matrix for different centrality classes are calculated assuming thermal production [Eq. (\ref{density})] and are presented as functions of the angle with respect to the impact parameter. The calculations are performed within a blast-wave framework incorporating elliptic flow and polarization effects \cite{florkblast,florkblast2}, together with a global vorticity as modeled in Ref. \cite{karpenko}. For quantitative comparisons with experimental data, blast-wave–type models such as that of Ref. \cite{ryb} are required.
   
As argued in \cite{kayman1} particles of higher spin than 1/2, carrying more than one qubit of information, are optimal probes of such non-equilibrium. This is because the angular distribution of decay products in the hadron rest frame, $W(\theta,\phi,\rho_{ij})$ (where $\theta$ is aligned with the spin and $\phi$ is perpendicular to it) is related to the density matrix elements by a simple partial wave analysis (see the next section for an example), allowing us direct access to {\em most} (or all, for weak decays) $\rho_{ij}$ matrix elements, including the diagonal ones.

\begin{figure*}\begin{center}
\epsfig{width=15cm,clip=,figure=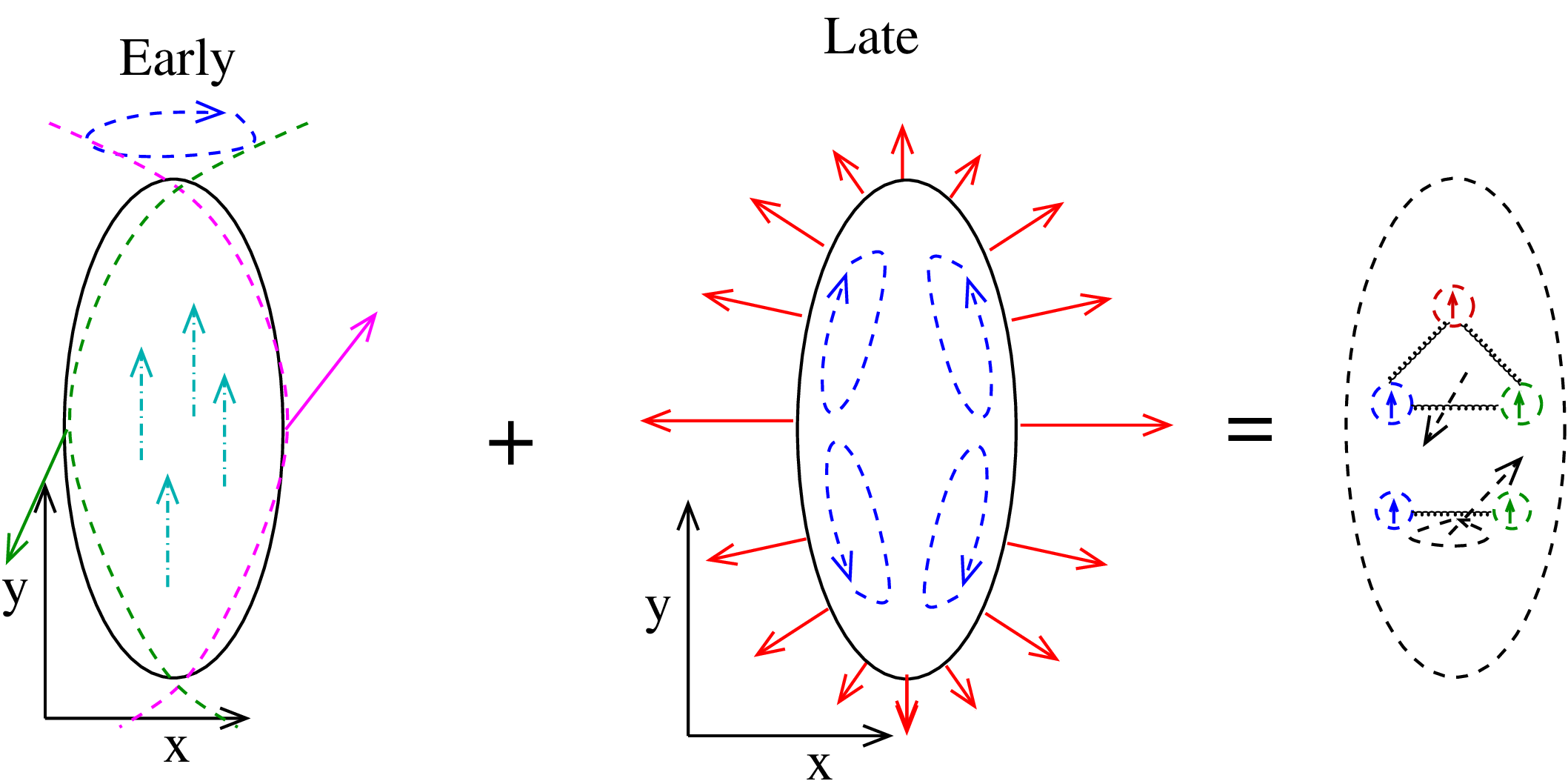}\caption{A schematic representation of the two vorticities in an event with finite impact parameter, together with the expected hadron structure at freeze-out from a coalescence type picture. The impact parameter and beam direction  are the $x$ and $z$ directions, respectively.  \label{fig2vort}}
\end{center}
\end{figure*}

Of course, the decay frame ($\theta,\phi,p_T^*$) is generally different from the lab frame ($\theta_r,\phi_r,p_T$). For non-relativistic particles, the difference in frames is simplified by the approximate invariance of the angles, so the final distribution in lab coordinates is simply a convolution.
%Dnsity matrix elements for various centrality classes are computed using the approach developed in \cite{ryb} assuming thermal production (\eqn{density}) convoluted with  a blast wave picture with elliptic flow and polarization (as in \cite{florkblast,florkblast2}) as well as a global vorticity (as in \cite{karpenko}), which add to a non-trivial angular momentum in the rest frame of the $f_2$.
\begin{eqnarray}
\label{convol}
\frac{dN}{d\Omega}&=&\int p_T dp_T d y \left. d\phi \left(\frac{dN_{\text{parent}}}{p_T'dp_T' dy'  d \phi'}\right)\right|_{y'=y,p_T'=p_T,\phi'=\phi}\nonumber\\
&&\times\frac{\partial\left( y,p_T,\phi\right)}{\partial \left(\theta_r,\phi_r,p\right)}W\left(\theta,\phi,\rho_{ij}(\theta_r,\phi_r)\right),
\end{eqnarray}
where  $W\left(\theta,\phi,\rho_{ij}(\theta_r,\phi_r)\right)$ will be calculated in the next section using a phenomenological Lagrangian and the helicity formalism.
For Bjorken type dynamics where coalescence is local in rapidity, the momentum and $\phi$ parts of the Jacobian are just normalization parameters so
\begin{equation}
\frac{\partial \left( \eta,p_T,\phi \right)}{\partial \left( \theta_r,\phi_r ,p\right)} \propto \frac{d \theta_r}{d\eta}  \eqcomma   \eta \simeq \tanh^{-1}\cos \left( \frac{\theta_r}{2} \right).
\end{equation}
Concretely, we know that there are two potential sources of vorticity common to all events: a ``global`` vorticity, which is common to the whole event and is formed at the initial collision due to the finite impact parameter, and ``local`` vorticity which is formed in concomitance with elliptic flow (Fig. \ref{fig2vort}). If $\tau_\Omega$ in \eqn{fertvalet} is indeed comparable to the lifetime of the fireball, then one expects spin to be more aligned in the global direction, while vorticity to be a sum of global and local directions (Fig \ref{fig2vort}). While a numerical calculation is needed for quantitative modeling, it is clear that this will give substantial differences between a density matrix of the form \eqn{density} and one of the form \eqn{coalmeson}.
%%%%%%%%%%%%%%%%%%%

In this work, we wish to concentrate on $f_2(1270)$, which is interesting in this regard for several reasons. It is a tensor ($S=2$) meson, whose density matrix is very large, 5$\times$5.
Even in a parity preserving decay we can obtain more information than spin alignment, since we can distinguish $\rho_{0,0}$ from $\rho_{\pm1,\pm1}$ and $\rho_{\pm1,\pm1}$ from $\rho_{\pm2,\pm2}$ as well as identify the non-diagonal matrix elements using the expressions for the whole spherical harmonic coefficients. \footnote{Perhaps a good name for such a measurement is spin projection} It should be noted that the $\pi-\pi$ invariant mass distribution of $f_2(1270)$ was recently measured by the ALICE collaboration using central-exclusive production events in pp collisions  \cite{Khatun:2024vgn}. Although the background conditions in heavy-ion collisions are expected to be significantly different, this measurement provides a valuable baseline and serves as a natural starting point for future studies in heavy-ion collision systems.

Moreover, according to the quark model, it is an $L=1$ angular momentum state, which, if the picture in \cite{kayman1} is correct, should be particularly affected by vorticity (similarly to quarkonium \cite{kayman2}).   

Furthermore, spin alignment {\em has} to be either aligned ($L=1$) or anti-aligned ($L=3$) for this particle to form. Thus, $P_L(\omega)$ here appears only as a normalization parameter (quarks in a misaligned spin and angular momentum state do not coalesce into $f_2$). Thus, both the abundance and the spin alignment of $f_2$ mesons, when compared to the abundance and polarization/alignment of other hadrons, is a promising probe of the dynamics between spin and vorticity.

For these reasons, the $f_2$ meson is a particularly ``clean`` probe for ascertaining if hadronization of spin is truly thermal (\eqn{density}), of coalescence type (\eqn{coalmeson}) or perhaps of another form. In the rest of this paper we will investigate this quantitatively.

This paper is organized as follows. In Section \ref{sec:f2decayangdist}, we evaluate the coupling strength between the $f_2$ and $\pi$ mesons in the $f_2\to\pi+\pi$(\textit{d-wave} decay) using an effective Lagrangian interaction. We then calculate the general angular distribution of the pions produced in the decay $f_2\to\pi+\pi$ using both the Lagrangian interaction formalism and the helicity formalism, and verify that the two methods yield identical results. In Section \ref{sec:blastwavemodel}, we introduce the parametrization of the elliptic fluid flow which accounts for the effects of the elliptic flow in the expanding medium. Furthermore, we use the anisotropic fluid flow parametrization to evaluate the fluid thermal vorticity. In Section \ref{sec:results}, we present the plot of the diagonal matrix element and $\rho_{20}$ of the spin density matrix of $f_2$ without and with global vorticity, as a function of the azimuthal angle in the $x-y$ plane for various classes of events centrality assuming thermal production.

\section{\label{sec:f2decayangdist}The decay width and angular distribution of $f_2\to\pi+\pi$}

To describe the decay $f_2(p)\to\pi(p_1)+\pi(p_2)$, we adopt the interaction Lagrangian introduced in Refs. \cite{Shklyar:2004ba,Suzuki:1993zs}.

\begin{align}
\mathcal{L}_{f_2\pi\pi}&=-\frac{2g_{f_{2}\pi\pi}}{m_{f_{2}}}\partial_{\mu}\boldsymbol{\pi}\vdot\partial_{\nu}\boldsymbol{\pi}f_{2}^{\mu\nu}.
\end{align}
Here, $f_2^{\mu\nu}$ denotes a symmetric massive spin-2 field. The invariant matrix element and the decay width are computed as below.

\begin{align}\label{eq:decaywidth}
\mathcal{M}&=\frac{g_{f_2\pi\pi}}{m_{f_2}}(p_1-p_2)_{\mu}(p_1-p_2)_{\nu}\varepsilon^{\mu\nu}(\boldsymbol{p},\lambda),\nonumber\\
\Gamma&=\frac{1}{8\pi}\frac{\abs{\boldsymbol{p}_1}}{m_{f_2}^{2}}\frac{1}{5}\frac{g_{f_2\pi\pi}^2}{m_{f_2}^2}\frac{32}{3}\abs{\boldsymbol{p}_1}^4=185.8\times(0.843)\;\text{MeV},\\
\abs{\boldsymbol{p}_1}&=\frac{1}{2}\sqrt{m_{f_2}^{2}-4m_{\pi}^{2}}=623\;\text{MeV}.\nonumber
%m_{f_2}&=1275\;\text{MeV},\;\;m_{\pi}=138.037\;\text{MeV}.\nonumber
\end{align}
$\lambda$ stands for the helicity of $f_2$. In $f_2$ rest frame where helicity is ill defined, it is defined to be a spin projection on $z$-axis, which depends on our choice of convention. The decay width, the branching ratio(number inside the parentheses) and the masses of the particles are adopted from \cite{ParticleDataGroup:2024cfk}. The dimensionless coupling constant is obtained to be $g_{f_{2}\pi\pi}=5.89$.

Now we compute the general angular distribution $W(\theta,\phi,\rho_{ij})$ of produced pions in $f_2\to\pi+\pi$ using phenomenological Lagrangian and helicity formalism in the rest frame of $f_2$ meson.

\subsection{Phenomenological Lagrangian}

Let us assume that initial $f_2$ is a superposition of different angular momentum states with respective amplitude $a_{\lambda}$. The matrix element of $f_2\to\pi+\pi$ in Eq. (\ref{eq:decaywidth}) is generalized as

\begin{align}\label{eq:generalf2}
\mathcal{M}&=\frac{4g_{f_{2}\pi\pi}}{m_{f_2}}p_{1\mu}p_{1\alpha}\sum_{\lambda=-2}^{2}a_{\lambda}\varepsilon^{\mu\alpha}(\boldsymbol{p},\lambda),\quad\rho_{\lambda\lambda^{\prime}}=a_{\lambda}a_{\lambda^{\prime}}^{*}.
\end{align}
We have used that $p_{\mu}\varepsilon^{\mu\alpha}(\boldsymbol{p},\lambda)=0$. $\rho_{\lambda\lambda^{\prime}}$ is the matrix element of a spin density operator. $W(\theta,\phi,\rho_{ij})$ can be obtained by taking a square of the matrix element in Eq. (\ref{eq:generalf2}), which is given in Appendix \ref{sec:spin2tensor}.

\begin{align}
W(\theta,\phi,\rho_{ij})&=\frac{1}{\Gamma}\frac{1}{32\pi^2}\frac{\abs{\boldsymbol{p}_1}}{m_{f_2}^2}\abs{\mathcal{M}}^2.
\end{align}
$\Gamma$ and $\abs{\boldsymbol{p}_1}$ are given in Eq. (\ref{eq:decaywidth}).
The result is same as given in Eq. (\ref{eq:f2dist}) obtained in the following subsection using the helicity formalism.

\subsection{Helicity formalism}

In 2-body decay, the initial particle is represented by an angular momentum eigenstate at rest $\ket{JM}$. Here $J$ and $M$ are total spin and its projection on the initial quantization axis. On the other hand, the final particle state is given as a simultaneous eigenstate of momentum and helicity, $\ket{\boldsymbol{p}s\lambda}$ in which $\boldsymbol{p}$, $s$ and $\lambda$ stand for the momentum, spin and helicity of a final particle, respectively. The final particle angular distribution is evaluated in the rest frame of the initial particle. Let the final particle momentum be written in terms of the spherical coordinate $(p,\theta,\phi)$, where $\theta$ is the polar angle between the final particle momentum and the initial quantization axis and $\phi$ is an azimuthal angle.

\begin{figure*}[t]
\centering
\includegraphics[scale=1.2]{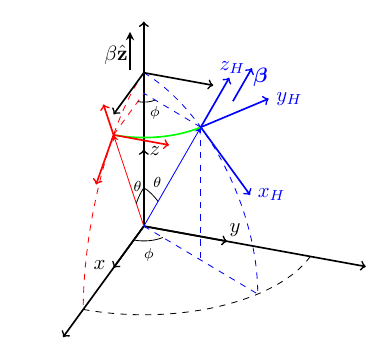}
\caption{\label{fig:helicityboost}Helicity Lorentz transformation of a particle. Subscript $H$ denotes the helicity rest frame.}
\end{figure*}

The helicity rest frame of a final particle is reached by a Lorentz boost along a $z$-axis, which is followed by a rotation parametrized by Euler angle of $(\phi,\theta,0)$. The helicty Lorentz transformation is illustrated in Fig. \ref{fig:helicityboost}.

We assume that the spin and helicity of the first and second final particles are $(s_1,\lambda_1)$ and $(s_2,\lambda_2)$, respectively. The formula for the normalized angular distribution of a final particle is given in \cite{Leader:2001nas}.
\begin{eqnarray}\label{eq:angdistformula}
W(\theta,\phi,\rho_{ij})&=&\sum_{\lambda_{1}\lambda_{2}}\sum_{MM^{\prime}}\frac{2J+1}{4\pi}\abs{F^J_{\lambda_1\lambda_2}}^{2}\nonumber\\
&&\times D^{J\dagger}\tensor{(\phi,\theta,0)}{^\lambda_M}\rho_{MM^{\prime}}D^{J}\tensor{(\phi,\theta,0)}{^{M^{\prime}}_\lambda},\\
\sum_{\lambda_1\lambda_2}\abs{F^J_{\lambda_1\lambda_2}}^2&=1&.\nonumber
\end{eqnarray}
Here, $\rho_{MM^{\prime}}$ stands for the spin-density matrix of the  initial particle. $D^J(\phi,\theta,0)$ is the Wigner D-matrix where $\lambda=\lambda_1-\lambda_2$ is the difference of helicity of final particles. $F^J_{\lambda_1\lambda_2}$ is the reduced helicity amplitude. For the decay $J$ into two spin-0 particles, the final particles satisfy $\lambda_1=\lambda_2=0$. In this case, the angular distribution $W(\theta,\phi,\rho_{ij})$ can be written in terms of the spherical harmonics as:
\begin{align}\label{eq:j00decay}
W(\theta,\phi,\rho_{ij})&=\sum_{MM^{\prime}}Y_J^M(\theta,\phi)Y_J^{M^{\prime}*}(\theta,\phi)\rho_{MM^{\prime}}.
\end{align}
The general angular distribution of produced pion is given in Appendix \ref{sec:twomethod}, Eq.(\ref{eq:f2dist}) as a function of $\theta$, $\phi$ and spin density matrix element of $f_2$ meson. After integrating over $\phi$, we obtain $W(\theta,\rho_{ii})$ as below.

\begin{widetext}

%\begin{eqnarray}W(\theta)&=&\frac{5}{2}\Bigg(\frac{\rho_{11}+\rho_{-1-1}+2(\rho_{22}+\rho_{-2-2})}{6}+\frac{\rho_{11}+\rho_{-1-1}-\rho_{22}-\rho_{-2-2}}{2}P_1(\cos\theta)^2\nonumber\\&&+\frac{\rho_{22}+\rho_{-2-2}-4(\rho_{11}+\rho_{-1-1})+6\rho_{00}}{6}P_2(\cos\theta)^2\Bigg).\end{eqnarray}

\begin{eqnarray}
\label{wonlydiag}
W(\theta,\rho_{ii})
%&=\frac{5}{2}\Bigg(\frac{1}{6}\left(\rho_{11}+\rho_{-1-1}+2(\rho_{22}+\rho_{-2-2})\right)+\frac{1}{2}(\rho_{11}+\rho_{-1-1}-\rho_{22}-\rho_{-2-2})P_1(\cos\theta)^2\nonumber\\
%&+\frac{1}{6}\left(\rho_{22}+\rho_{-2-2}-4(\rho_{11}+\rho_{-1-1})+6\rho_{00}\right)P_2(\cos\theta)^2\Bigg)\nonumber\\
&=&\frac{5}{16}\Bigg(2\rho_{00}+3(\rho_{22}+\rho_{-2-2})+2\left(-6\rho_{00}+6(\rho_{11}+\rho_{-1-1})-3(\rho_{22}+\rho_{-2-2})\right)\cos^2\theta\nonumber\\
&&+3\left(6\rho_{00}-4(\rho_{11}+\rho_{-1-1})+\rho_{22}+\rho_{-2-2}\right)\cos^4\theta\Bigg).
\end{eqnarray}
Note, $\rho_{22}+\rho_{11}+\rho_{00}+\rho_{-1-1}+\rho_{-2-2}=1$.
%\begin{align}W(\theta)&=\frac{5}{2}\Bigg(\frac{\rho_{11}+\rho_{-1-1}+2(\rho_{22}+\rho_{-2-2})}{6}+\frac{\rho_{11}+\rho_{-1-1}-\rho_{22}-\rho_{-2-2}}{2}P_1(\cos\theta)^2+\frac{\rho_{22}+\rho_{-2-2}-4(\rho_{11}+\rho_{-1-1})+6\rho_{00}}{6}P_2(\cos\theta)^2\Bigg).\end{align}

\end{widetext}

\section{\label{sec:blastwavemodel}The blast wave model}

We proceed with our calculation of the spin density matrix element of $f_2$ in the medium created in heavy-ion collisions, where we can neglect baryon number density. To describe the medium we will use a simplified model, developed in \cite{florkblast,florkblast2} in which the transverse component of the fluid follows an elliptical expansion, while the longitudinal component is assumed to be boost-invariant. The elliptical parameterization of the region created in this interaction in the $x-y$ plane allows us to consider the effects of elliptic flow, which is described by the following expressions \cite{florkblast}:

\begin{equation}
    \begin{array}{c}
    x = r_{\text{max}}\sqrt{1-\epsilon}\cos\phi_r,\\
    y = r_{\text{max}}\sqrt{1+\epsilon}\sin\phi_r,
    \end{array}
    \label{transflow}
\end{equation}
where $\epsilon$(spatial size deformation) and $r_{\text{max}}$(transverse size) are model parameters and $\phi_r$ is azimuthal angle in the transverse plane. The flow velocity is defined by:
\begin{equation}
    u^{\mu} = \left(\frac{t}{N},\frac{x}{N}\sqrt{1+\delta},\frac{y}{N}\sqrt{1-\delta},\frac{z}{N}\right),
    \label{flow}
\end{equation}
$\delta$ is the parameter that accounts for the anisotropy of the transverse flow, and $N$ is a normalization parameter given by:
\begin{equation}
    N = \sqrt{\tau^2-(x^2-y^2)\delta} \eqcomma \tau^2 = t^2-x^2-y^2-z^2,
\end{equation}
where the spatial component $z$ is
\begin{equation}
	z = \sqrt{\tau^2_f+x^2+y^2}\sinh\eta.
        \label{para2}
\end{equation}
$\eta$ is space-time rapidity and $\tau_f$ is lifetime of a system. At this point, we can determine the thermal vorticity:
\begin{equation} 
    \Omega_{\mu,\nu} = -\frac{1}{2T}(\partial_{\mu} u_\nu - \partial_\nu u_\mu) - \frac{1}{2T^2}(u_\mu\partial_\nu T - u_\nu\partial_\mu T).
    \label{bwvor}
\end{equation}
It was shown in \cite{florkblast} that the first and second terms on the right-hand side of Eq.~(\ref{bwvor}) are equal, using the relativistic hydrodynamic equations at zero baryon density. The spatial components of the thermal vorticity, where $i, j, k = 1, 2, 3$, are defined as:
\begin{equation}
    \Omega_i = \frac{1}{2}\epsilon_{ijk}\Omega_{jk},
    \label{spacial_vorticity}
\end{equation}
where its values are given explicitly:
\begin{align}\label{vort}
\Omega_1&=\frac{yz}{TN^3}\left(1-\delta-\sqrt{1-\delta}\right),\nonumber\\
\Omega_2&=\frac{zx}{TN^3}\left(\sqrt{1+\delta}-1-\delta\right),\\
\Omega_3&=\frac{xy\sqrt{1-\delta^2}}{TN^3}\left(\sqrt{1+\delta}-\sqrt{1-\delta}\right).\nonumber
\end{align}
While the model in \cite{ryb} does not have global polarization (\cite{ryb} added it at the {\em quark level}), it can easily be added by
\begin{equation}
\Omega_2(x,y,z)\rightarrow \Omega_2(x,y,z) +\Omega_{\text{global}}.
\end{equation}
where $\Omega_{\text{global}}$ is a constant throughout the event, whose numerical estimate is approximately given in the introduction of \cite{karpenko} as $0.02\;\text{fm}^{-1}\rightarrow 0.00394 \;\text{GeV}$, therefore we have $\Omega_{\text{global}}\rightarrow\Omega_{\text{global}}/T = 0.024$ where $T=0.165\;\text{GeV}\rightarrow 0.837 \;\text{fm}^{-1}$.

\section{Results and discussion}\label{sec:results}

\begin{figure*}[t]
		\centering
    \epsfig{width=0.49\textwidth,figure=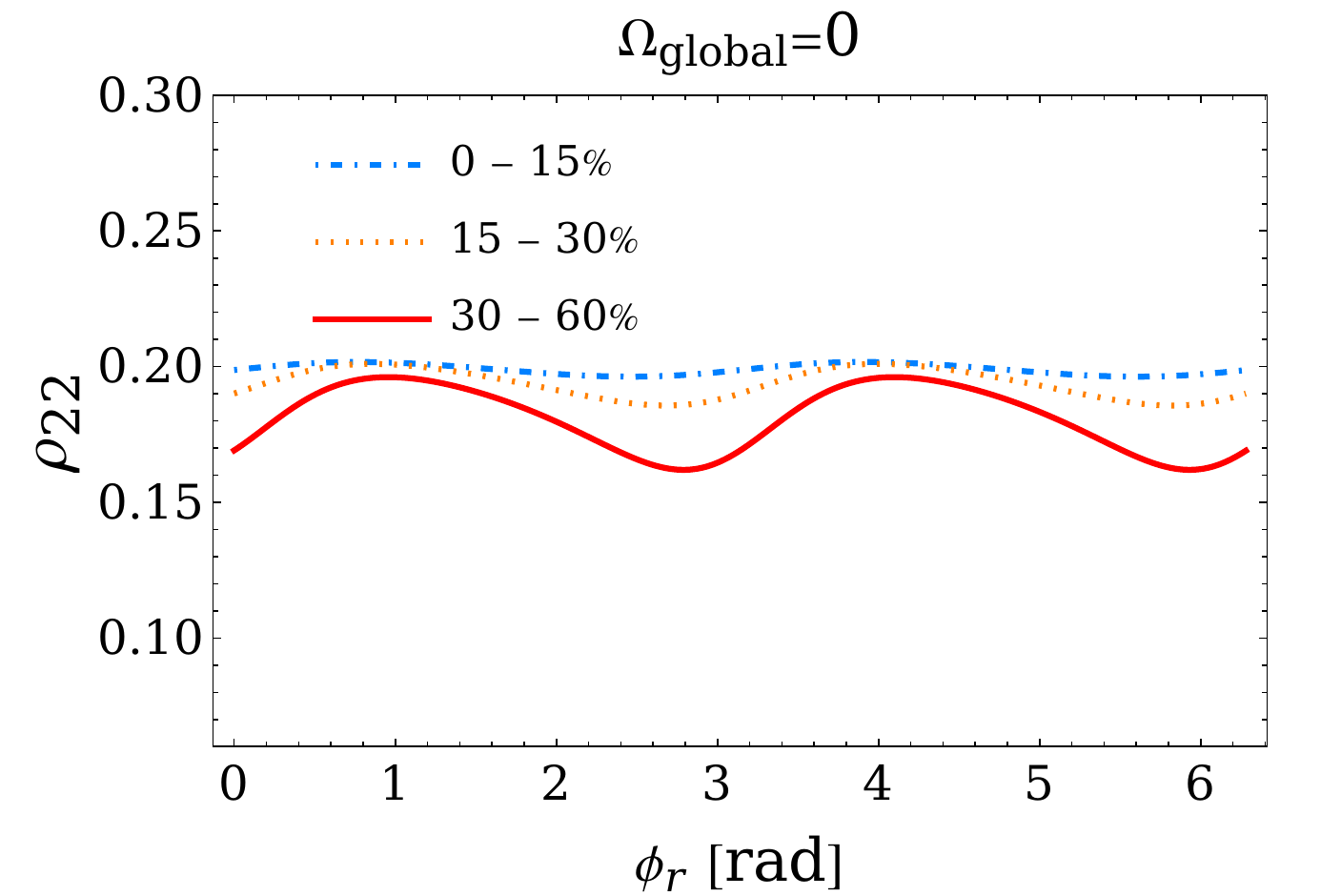}
   \epsfig{width=0.49\textwidth,figure=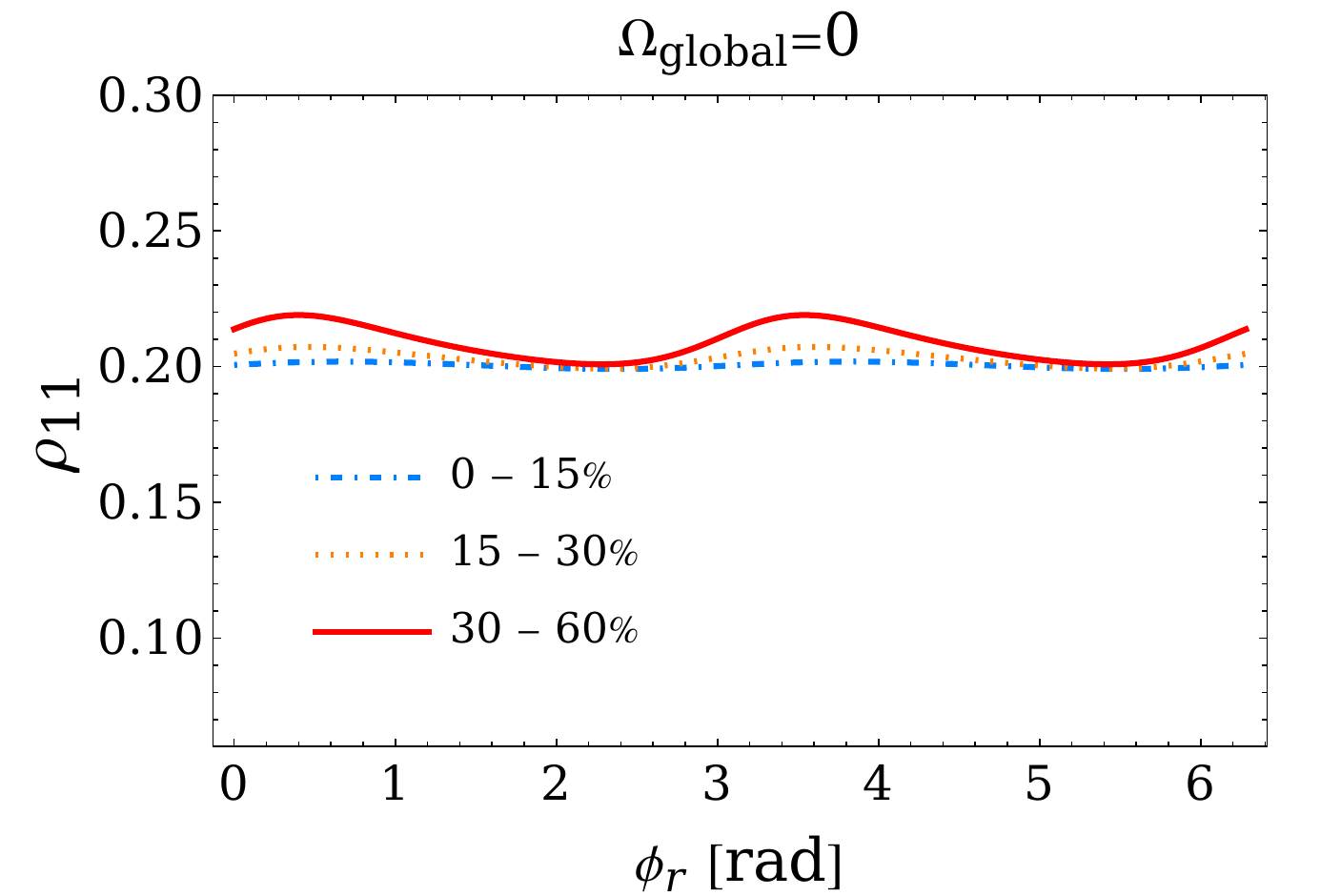}
   \epsfig{width=0.49\textwidth,figure=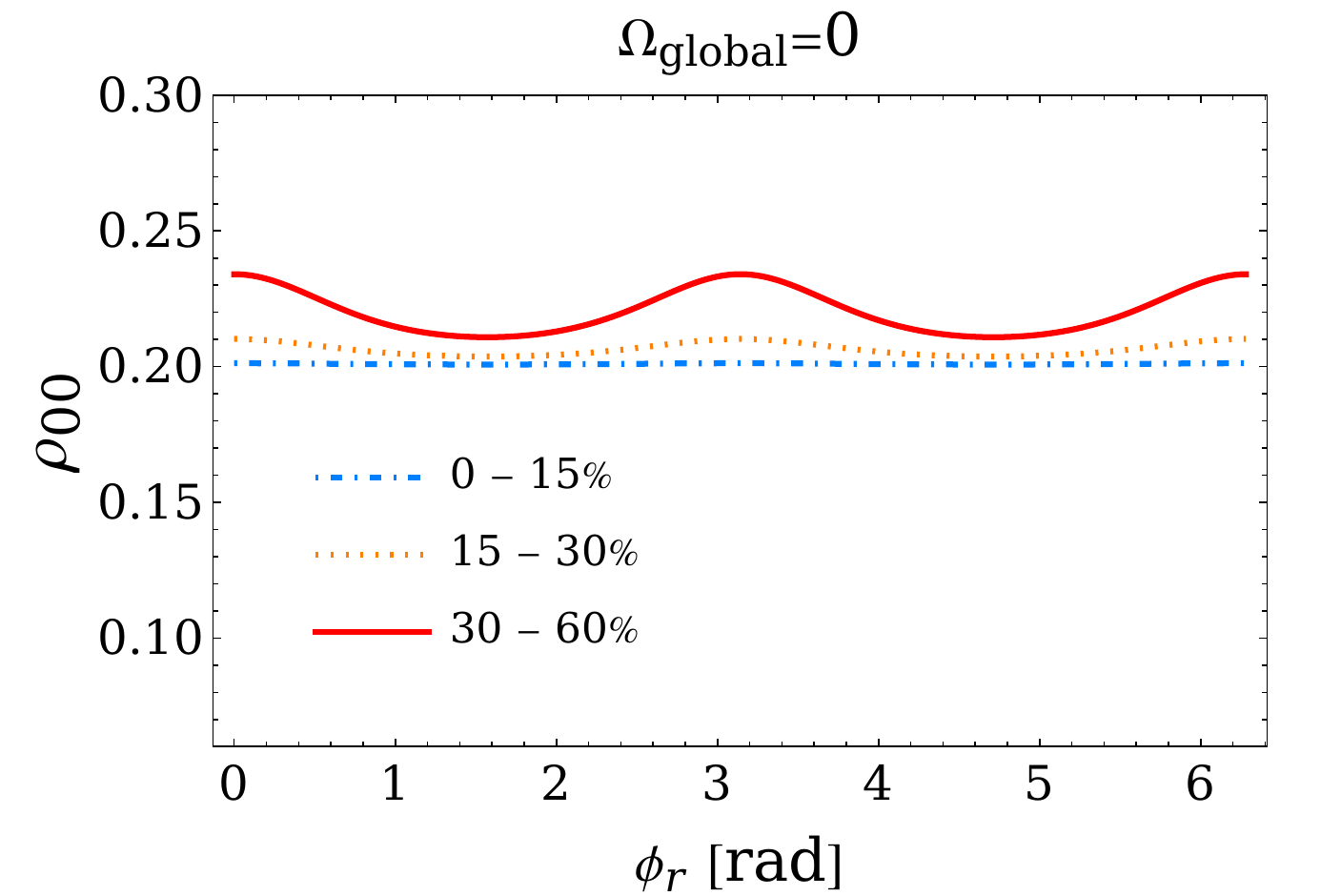}
   \epsfig{width=0.49\textwidth,figure=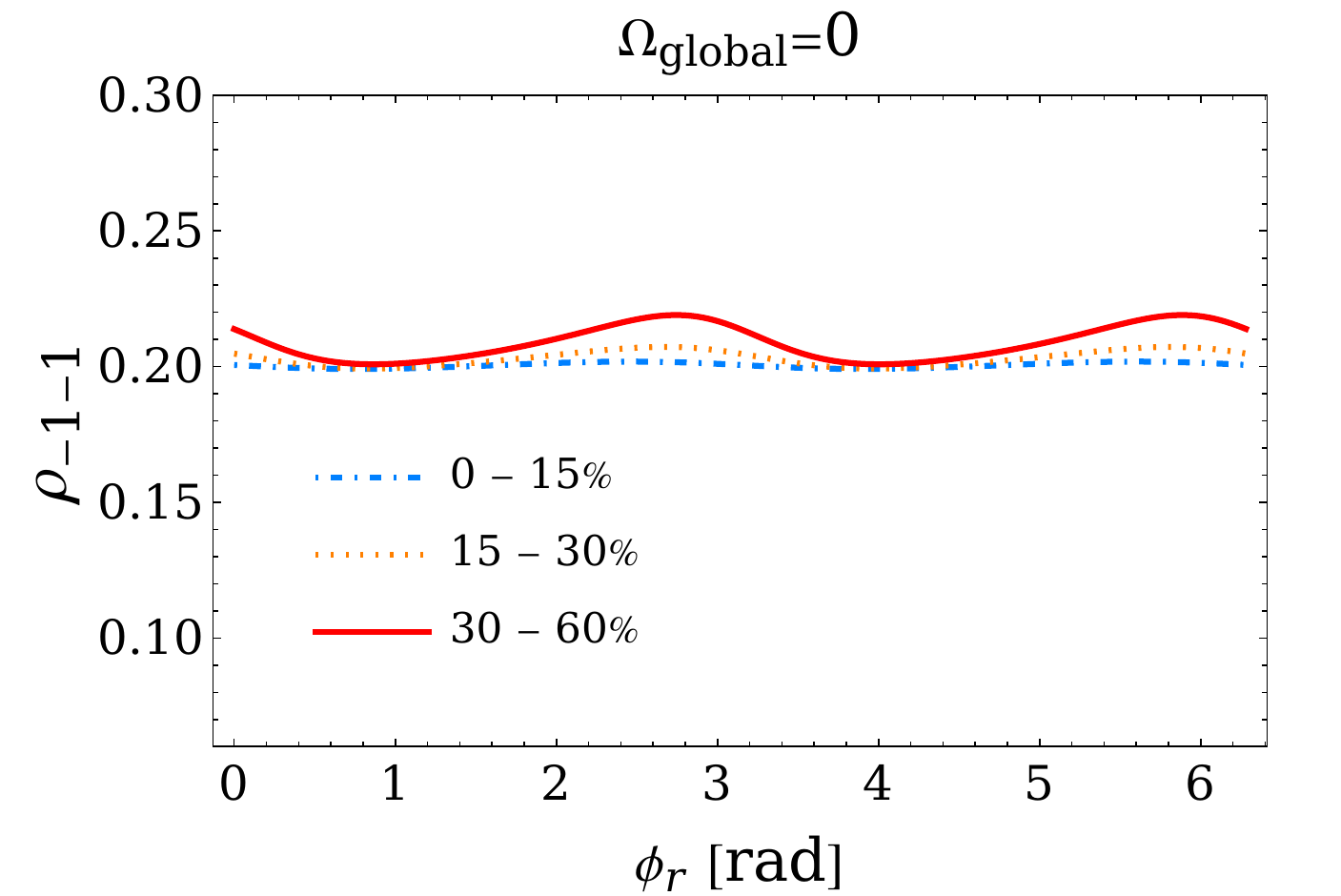}
   \epsfig{width=0.49\textwidth,figure=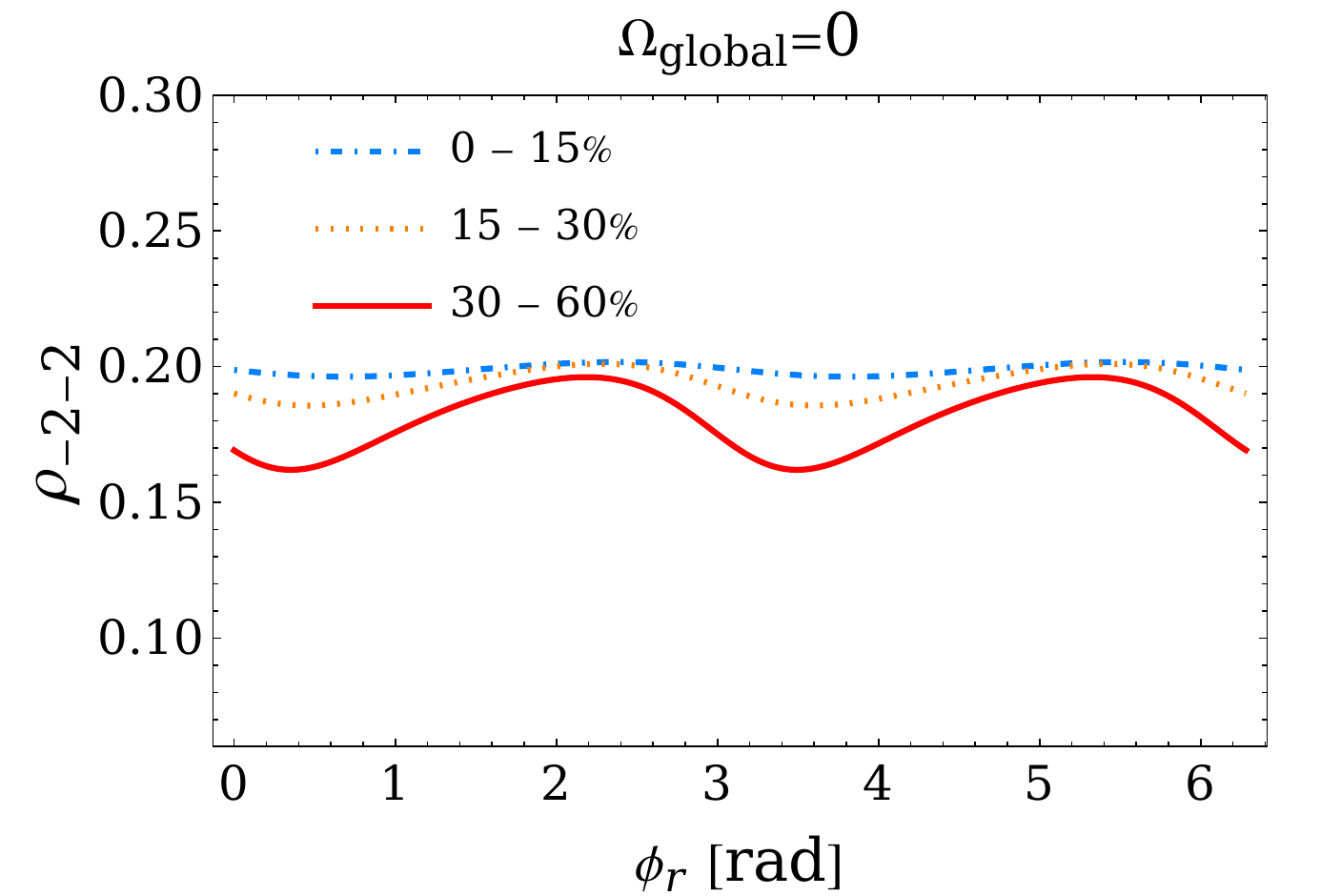}
            \caption{ Diagonal density matrix elements for various event centrality classes computed assuming thermal production (\eqn{density}) as a function of the angle w.r.t. impact parameter, assuming the blast wave picture with elliptic flow and polarization , \cite{florkblast,florkblast2} and no global vorticity \label{fig:rho_coefficients_noglobal}.}
\end{figure*}

\begin{figure*}[t]
		\centering
    \epsfig{width=0.49\textwidth,figure=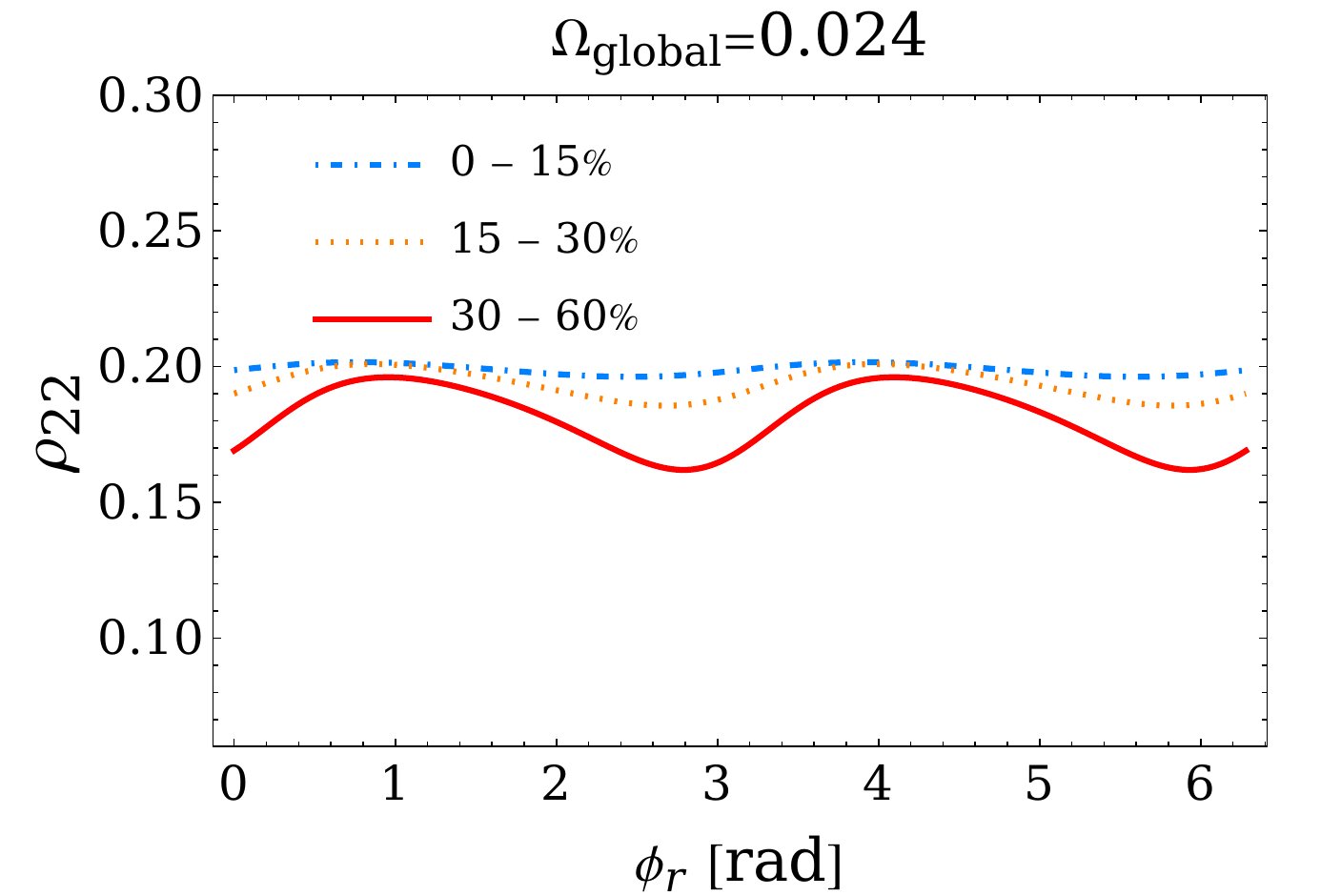}
   \epsfig{width=0.49\textwidth,figure=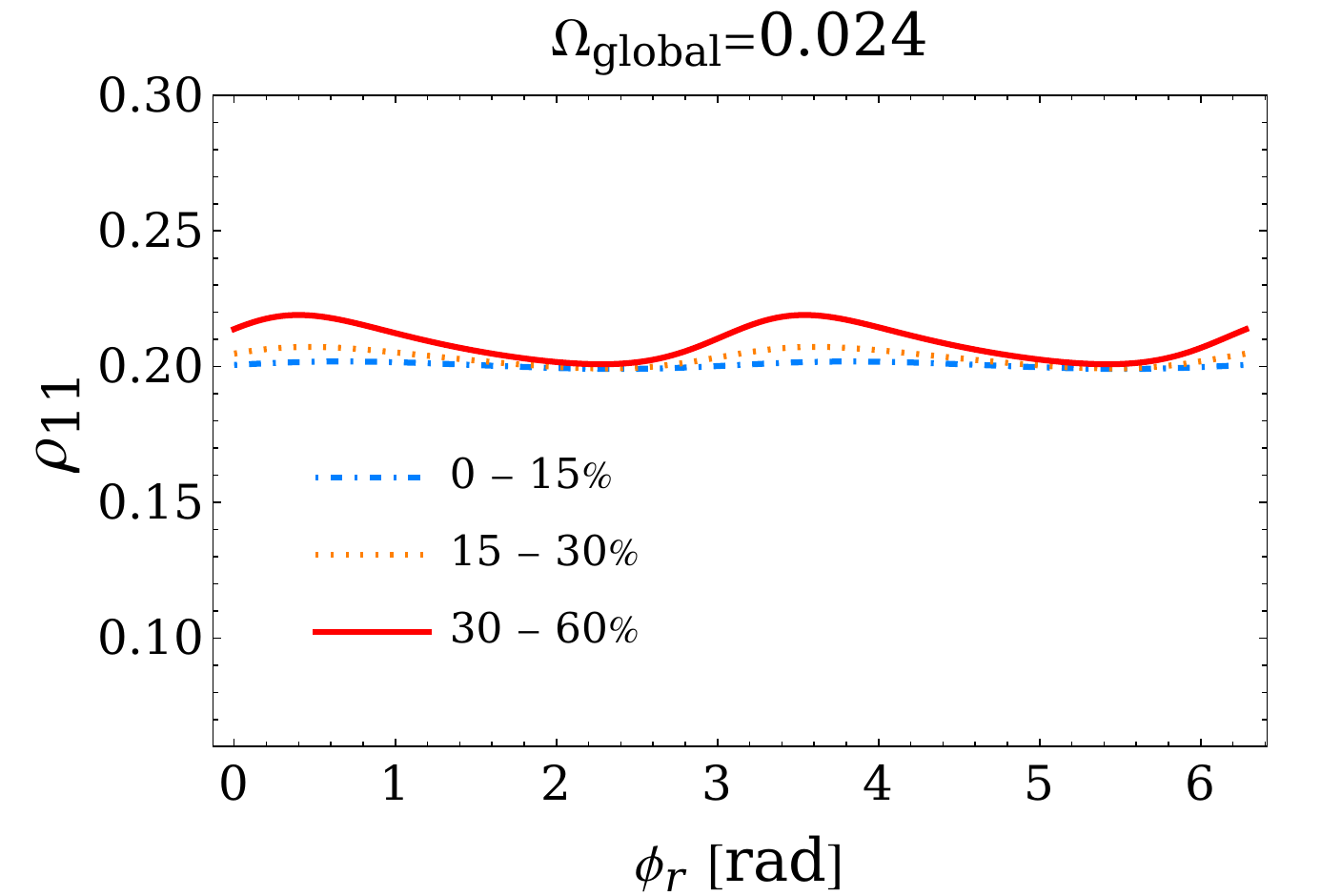}
   \epsfig{width=0.49\textwidth,figure=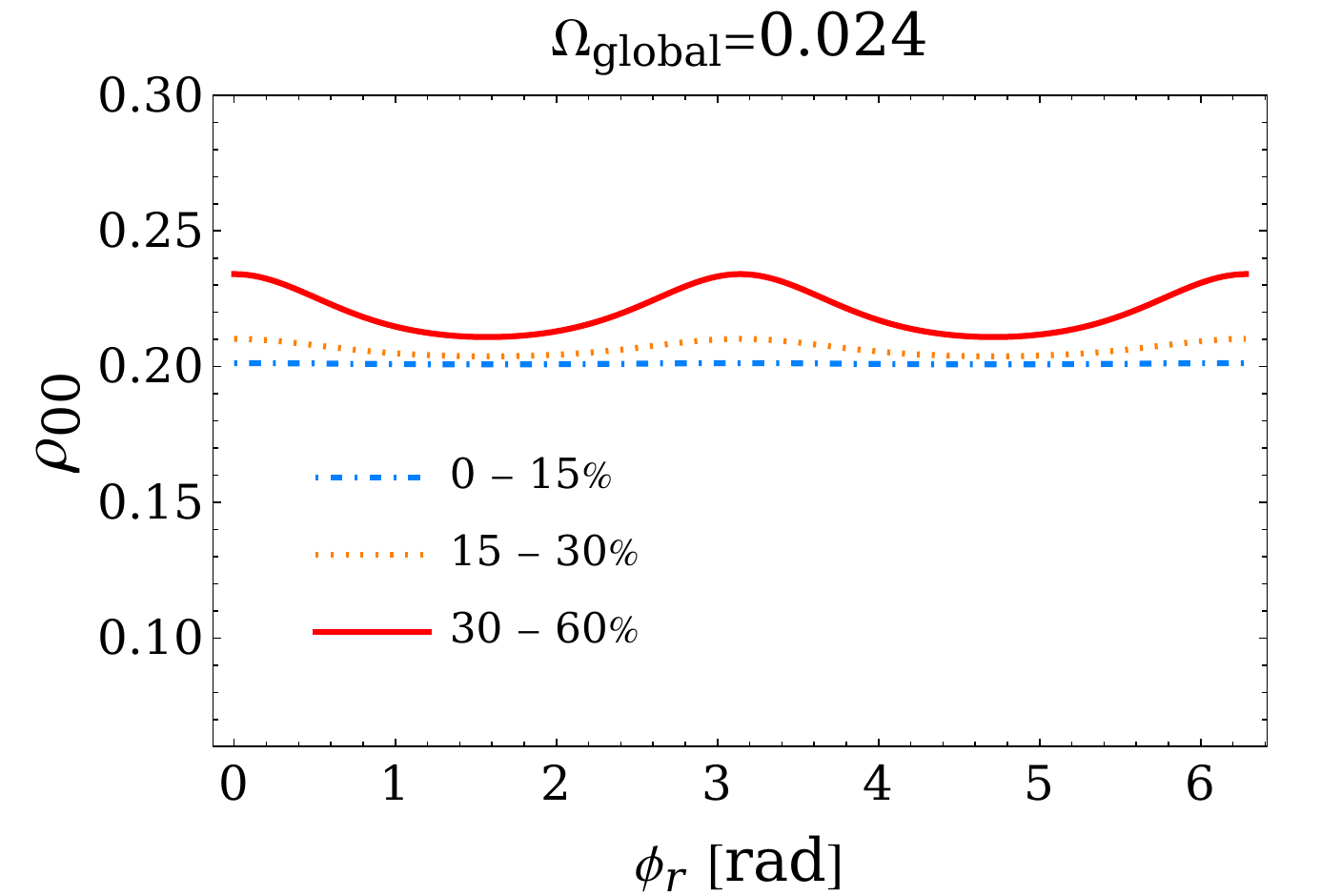}
   \epsfig{width=0.49\textwidth,figure=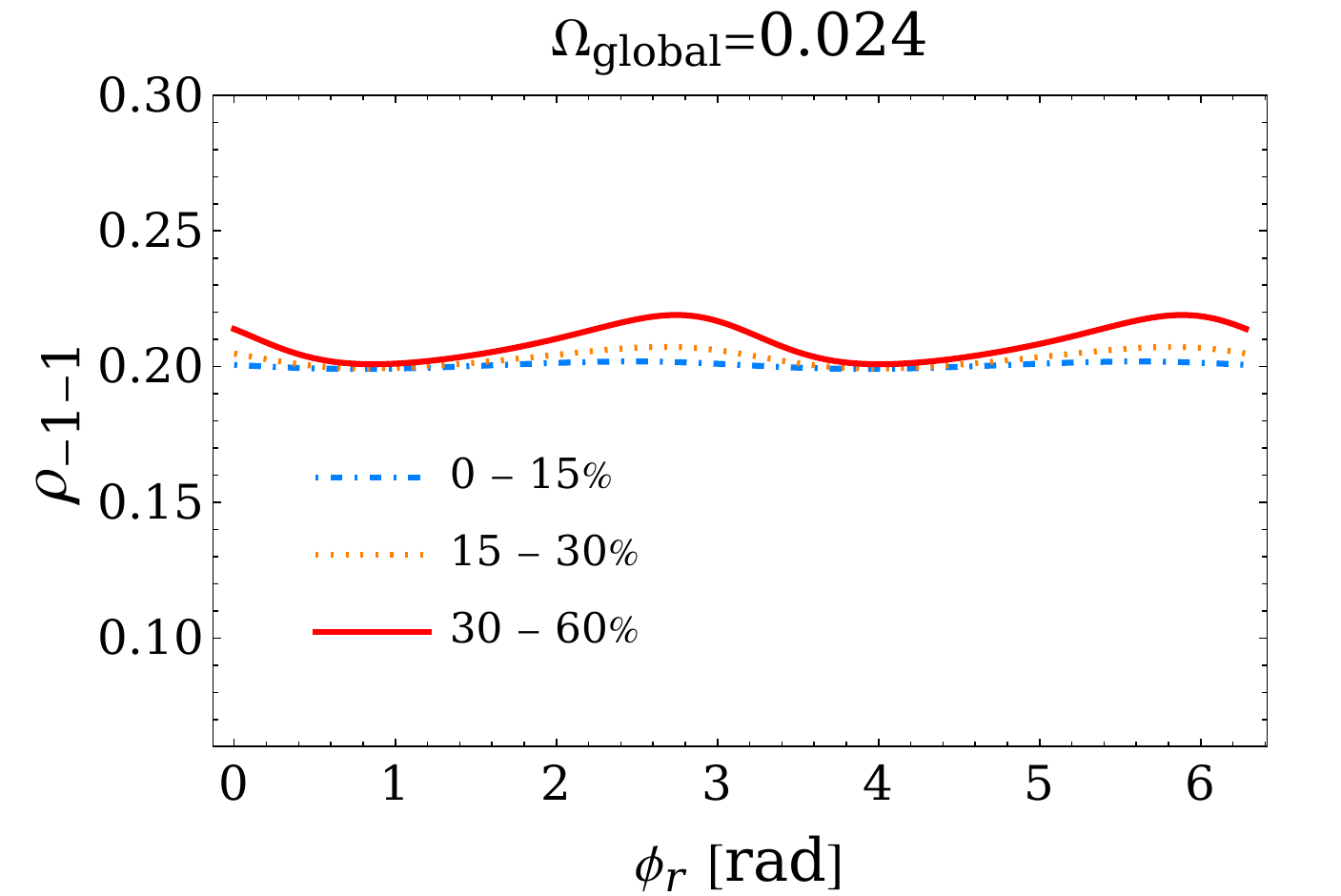}
   \epsfig{width=0.49\textwidth,figure=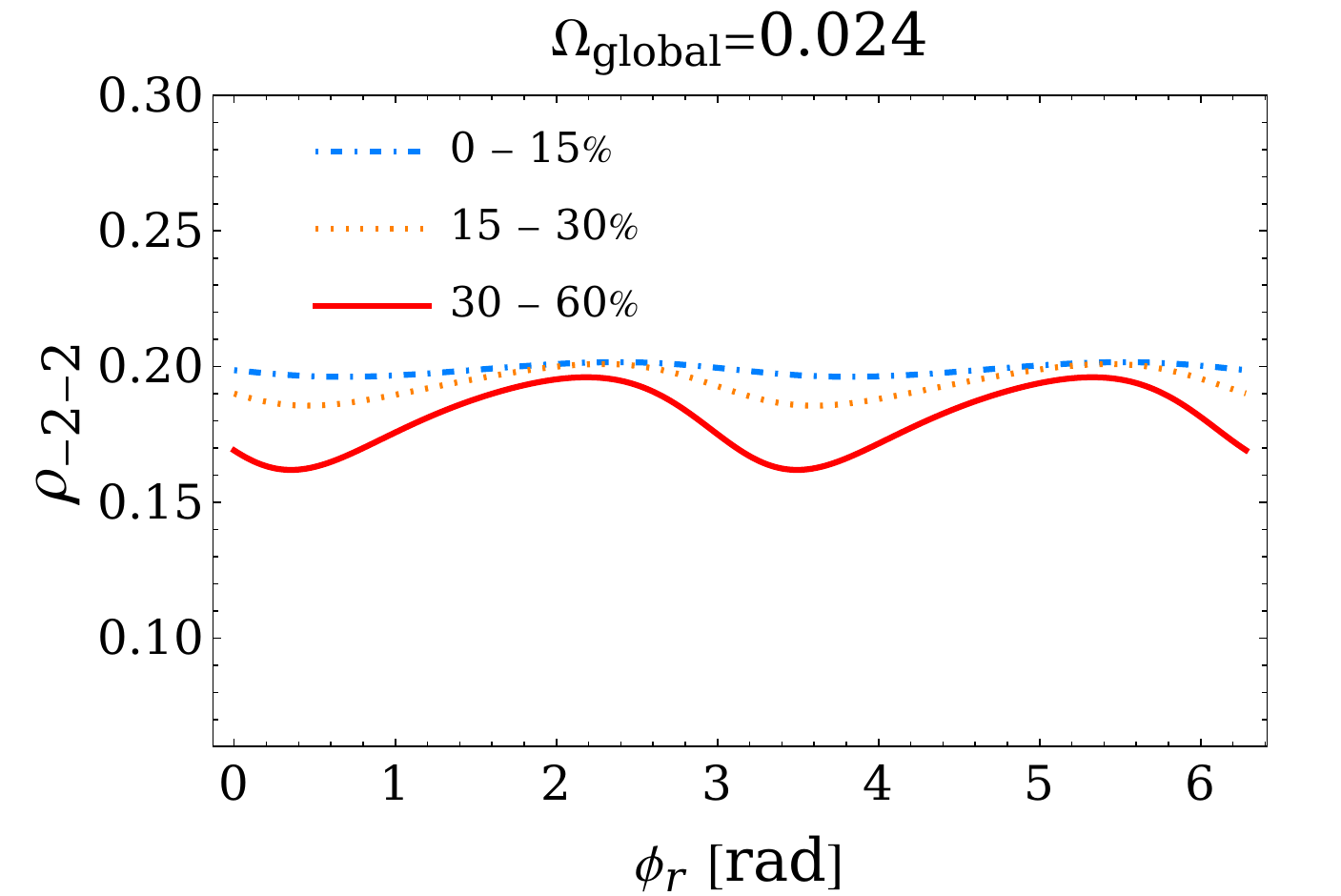}
            \caption{ Diagonal density matrix elements for various centrality classes, computed assuming thermal production (\eqn{density}) as a function of the angle w.r.t. impact parameter, assuming the blast wave picture with elliptic flow and polarization , \cite{florkblast,florkblast2} as well as a global vorticity of \cite{karpenko}\label{fig:rho_coefficients}.}
        
\end{figure*}

\begin{figure*}[t]
		\centering
    \epsfig{width=0.49\textwidth,figure=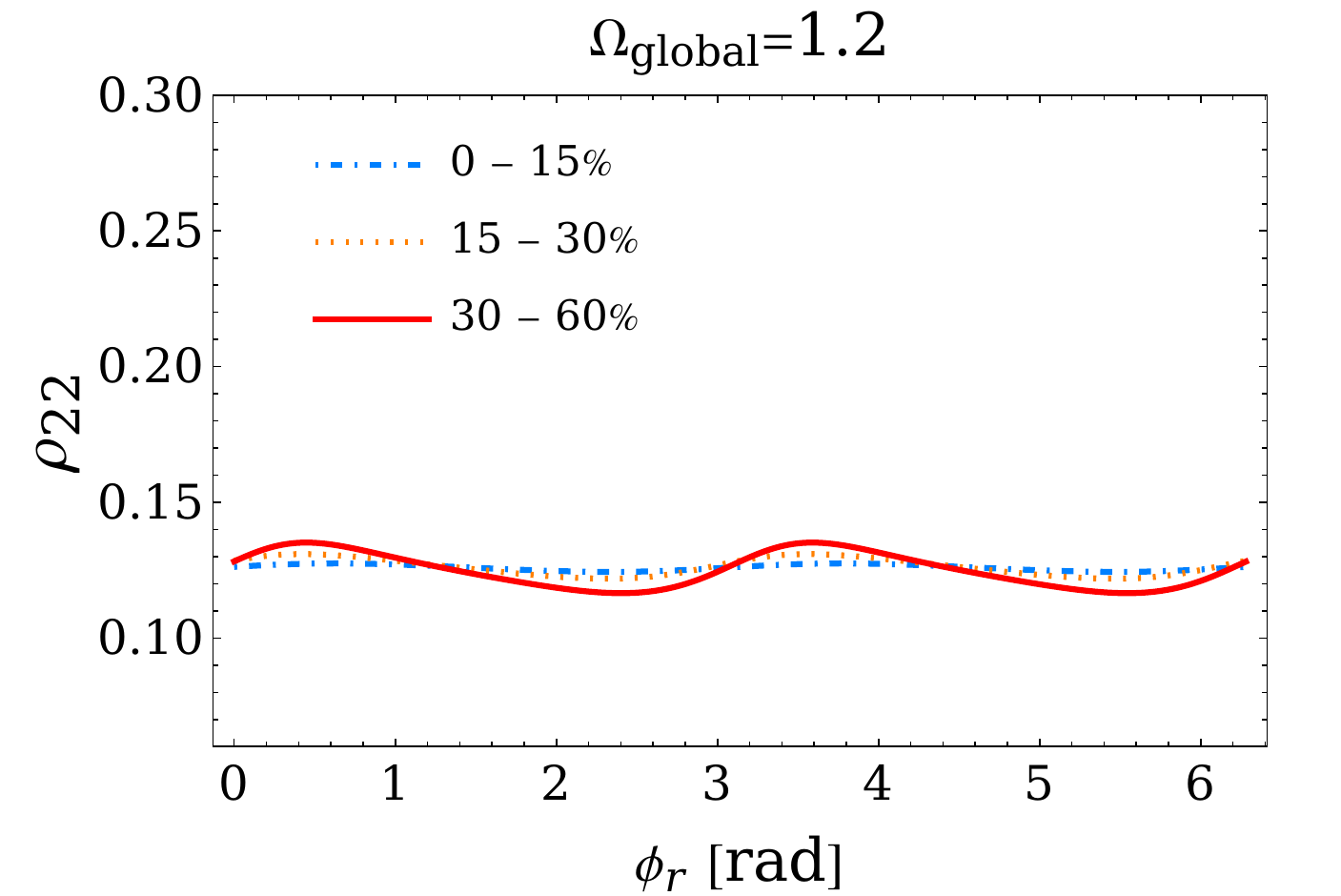}
   \epsfig{width=0.49\textwidth,figure=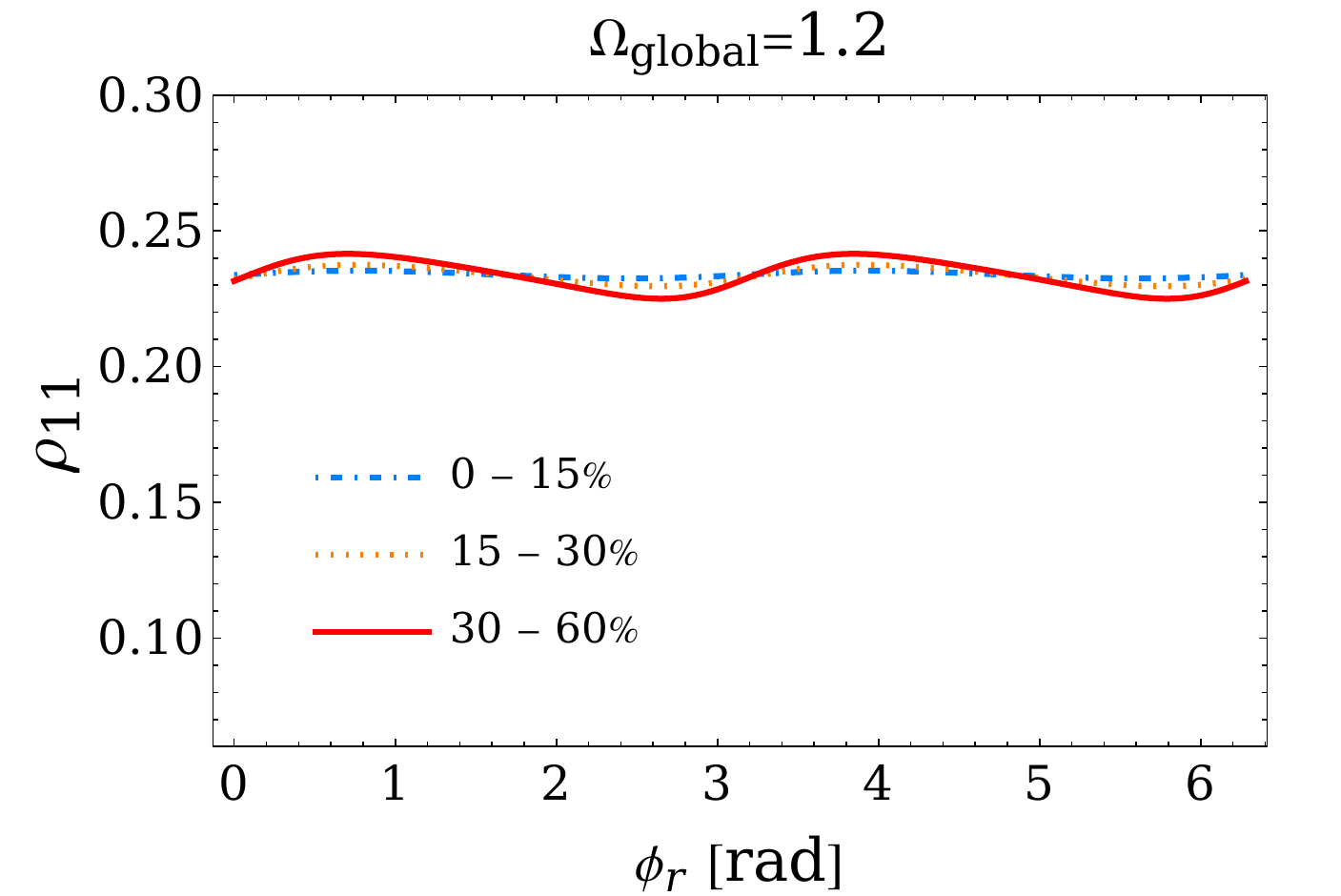}
   \epsfig{width=0.49\textwidth,figure=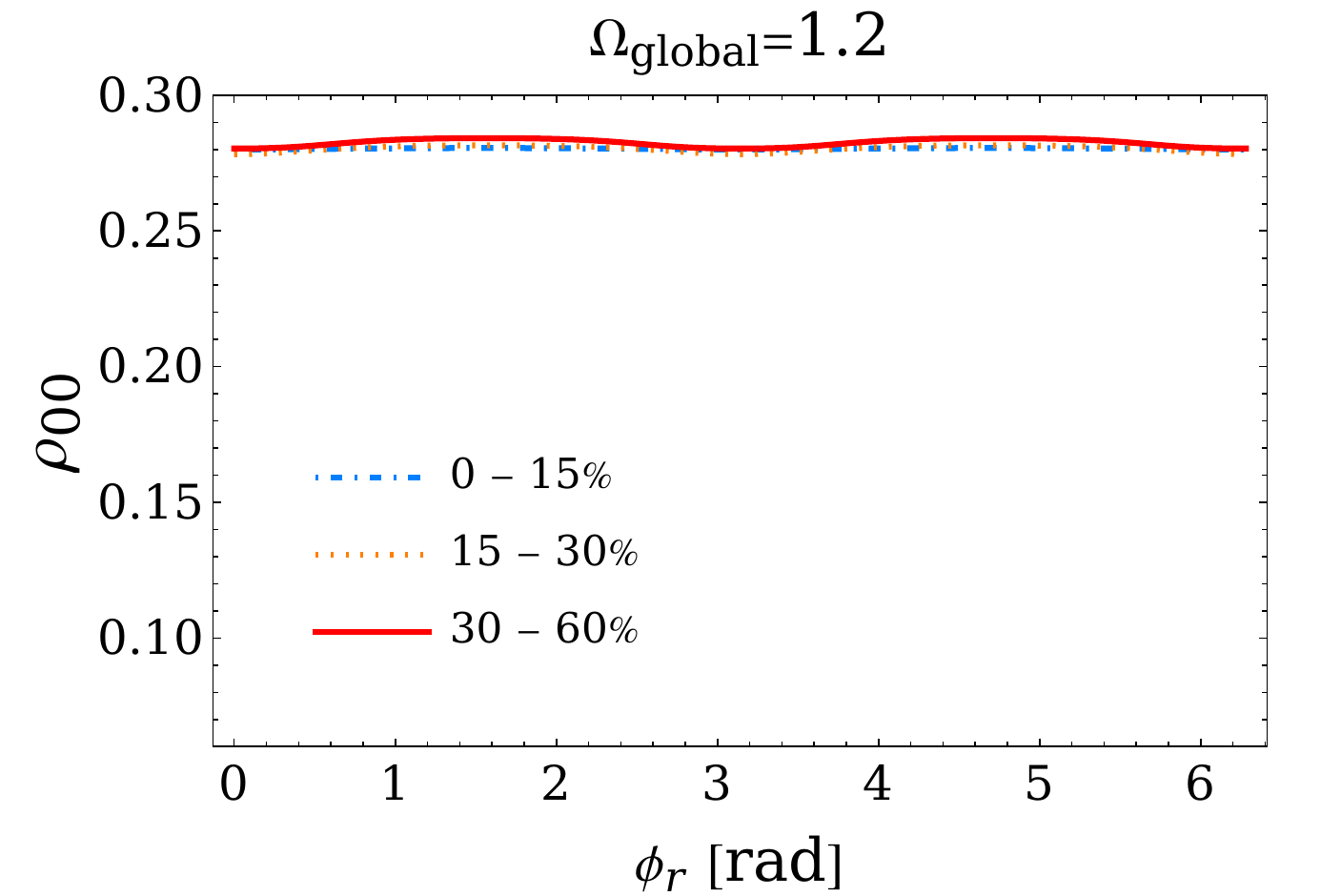}
   \epsfig{width=0.49\textwidth,figure=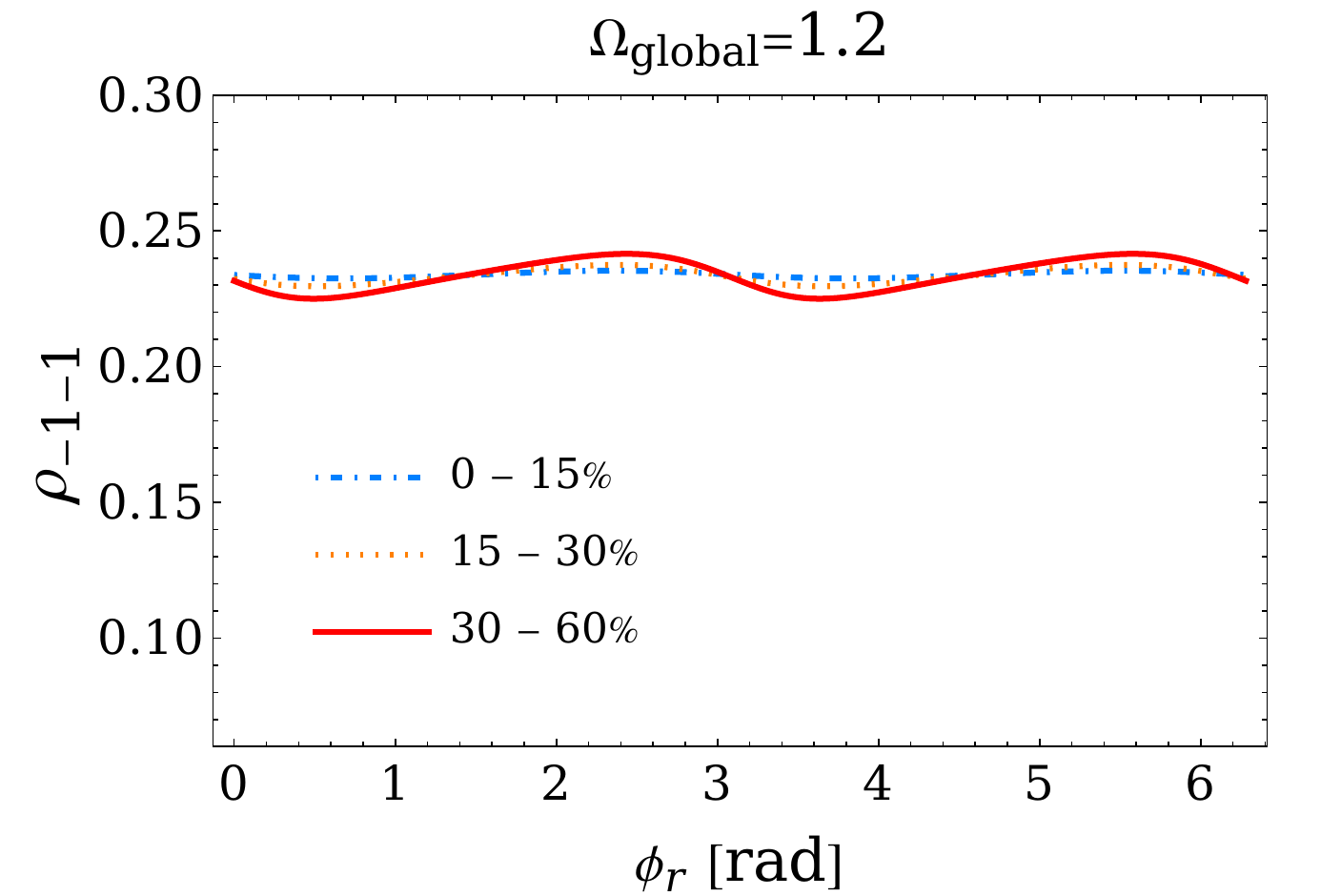}
   \epsfig{width=0.49\textwidth,figure=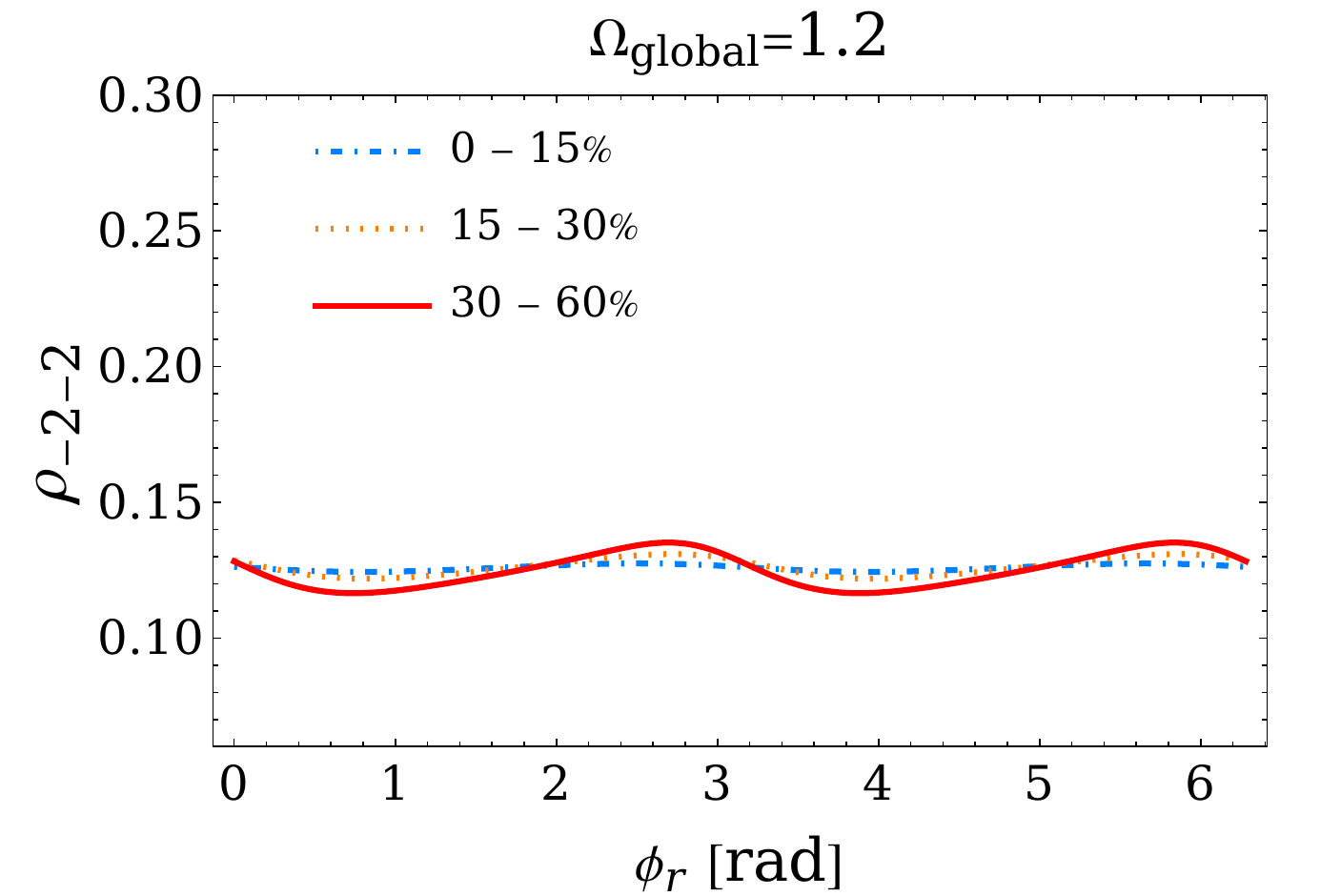}
            \caption{ Diagonal density matrix elements for various centrality classes, computed assuming thermal production (\eqn{density}) as a function of the angle w.r.t. impact parameter, assuming the blast wave picture with elliptic flow and polarization , \cite{florkblast,florkblast2} as well as a global vorticity of \cite{karpenko} multiplied by a factor of 50 \label{fig:rho_coefficients_largeomega}.}
        
\end{figure*}

\begin{figure*}[t]
		\centering
    \epsfig{width=0.49\textwidth,figure=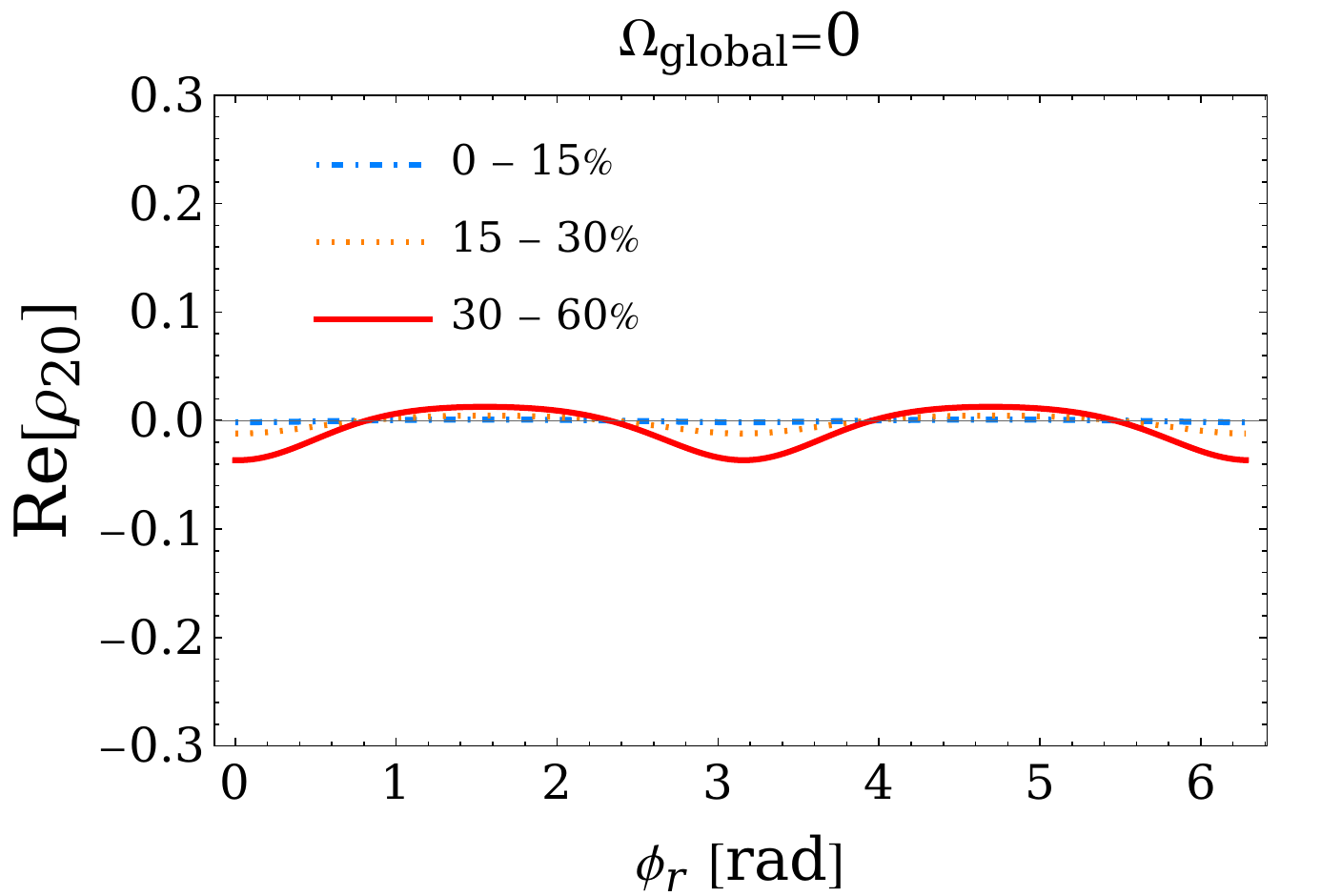}
   \epsfig{width=0.49\textwidth,figure=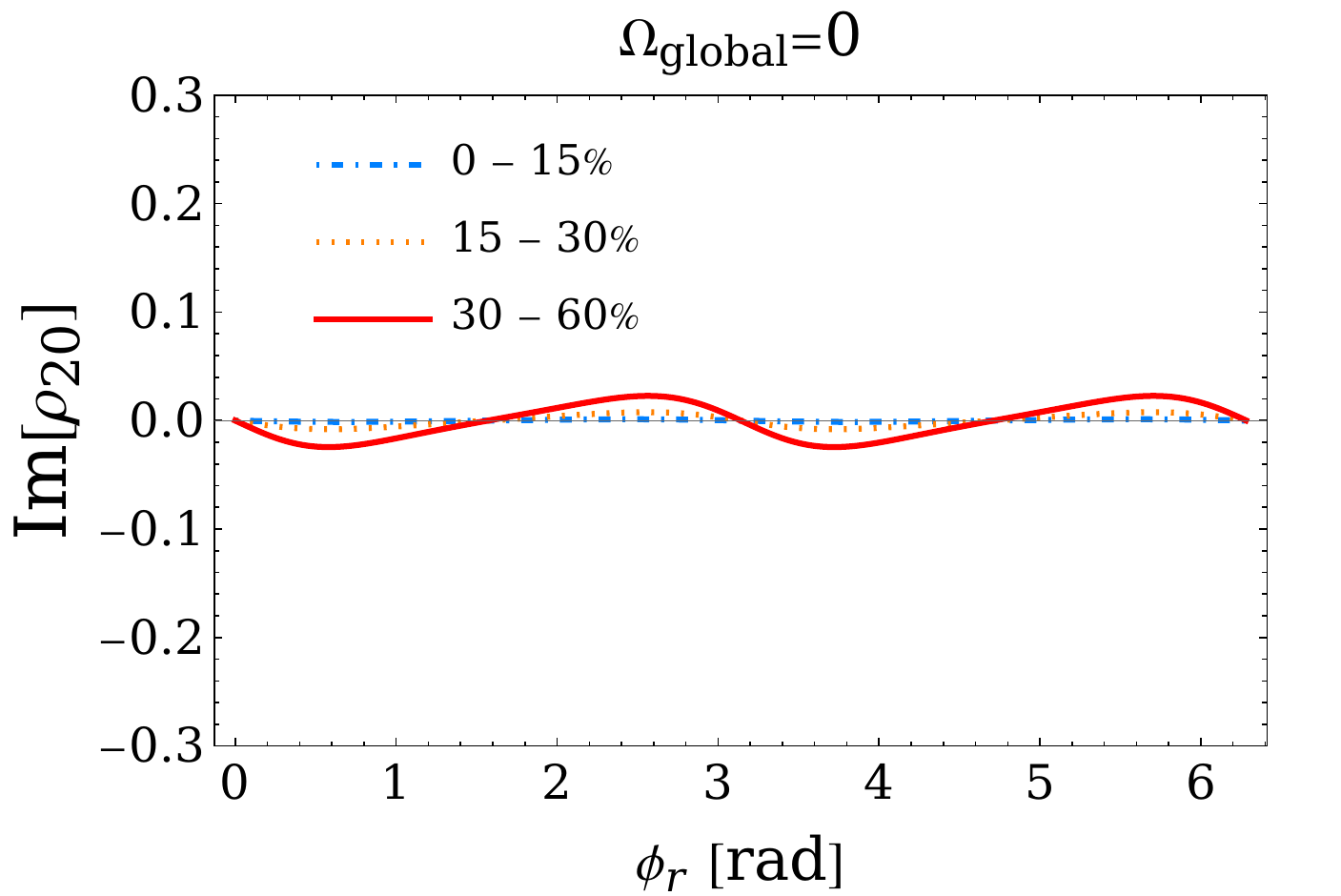}
  \epsfig{width=0.49\textwidth,figure=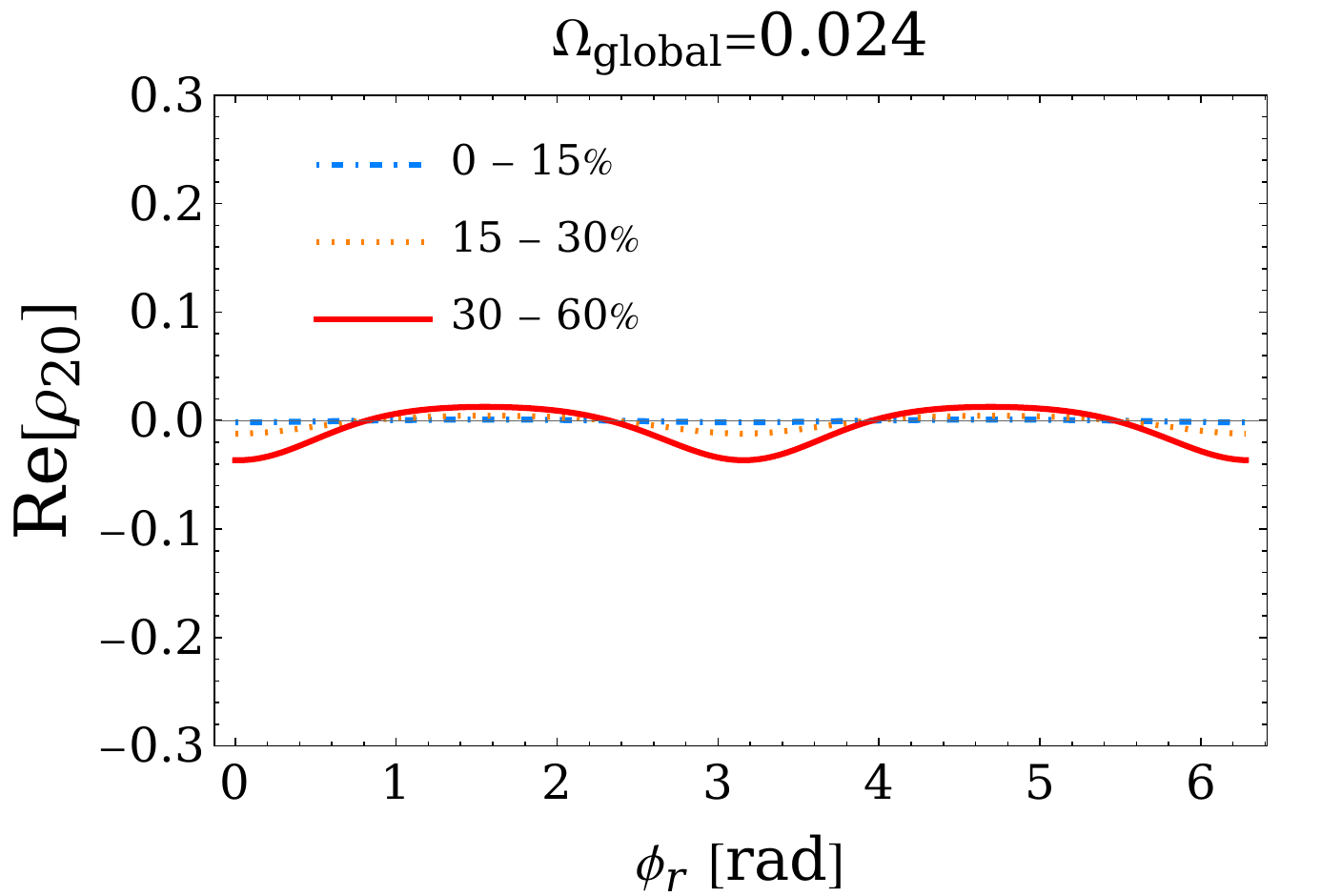}
   \epsfig{width=0.49\textwidth,figure=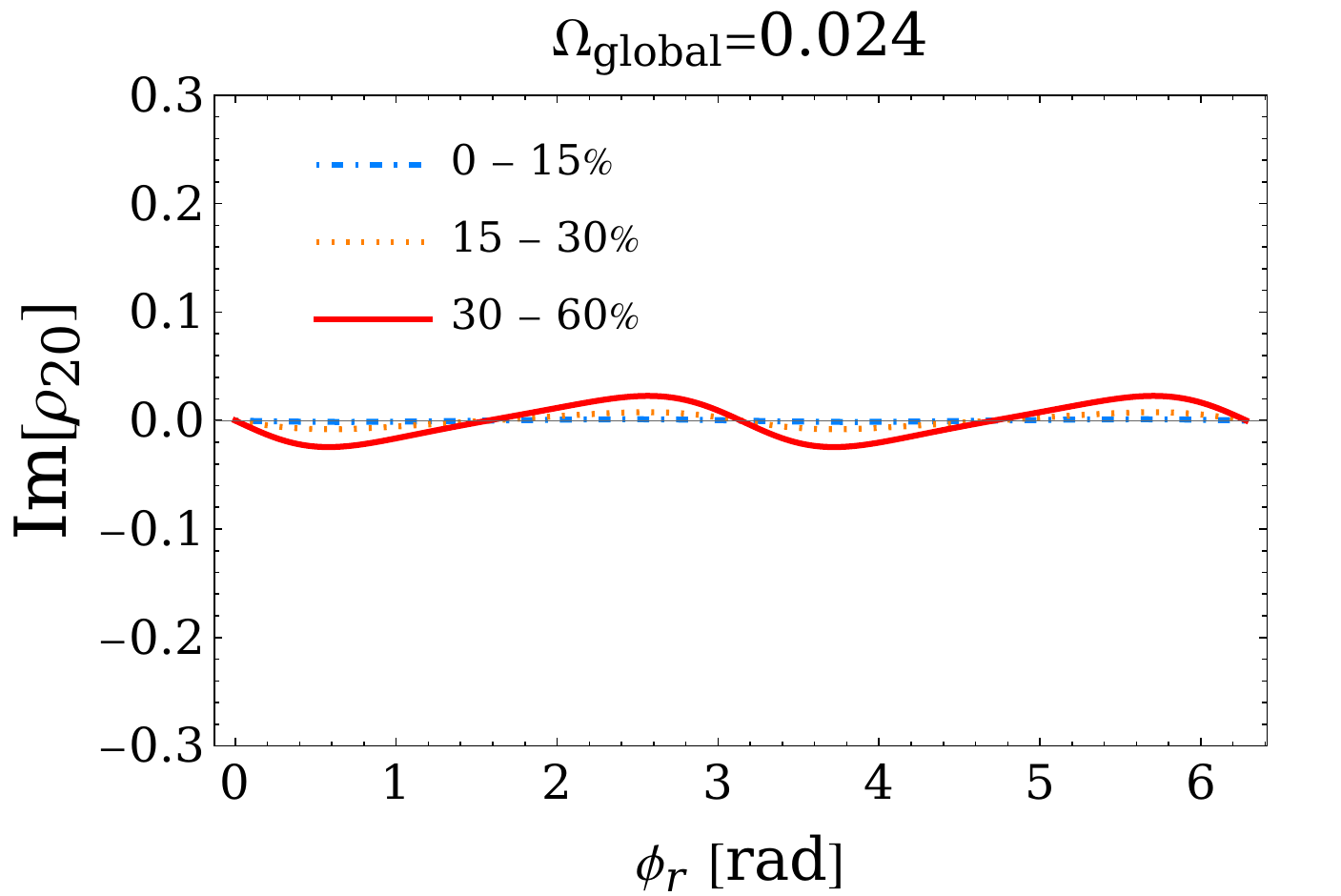} 
   \epsfig{width=0.49\textwidth,figure=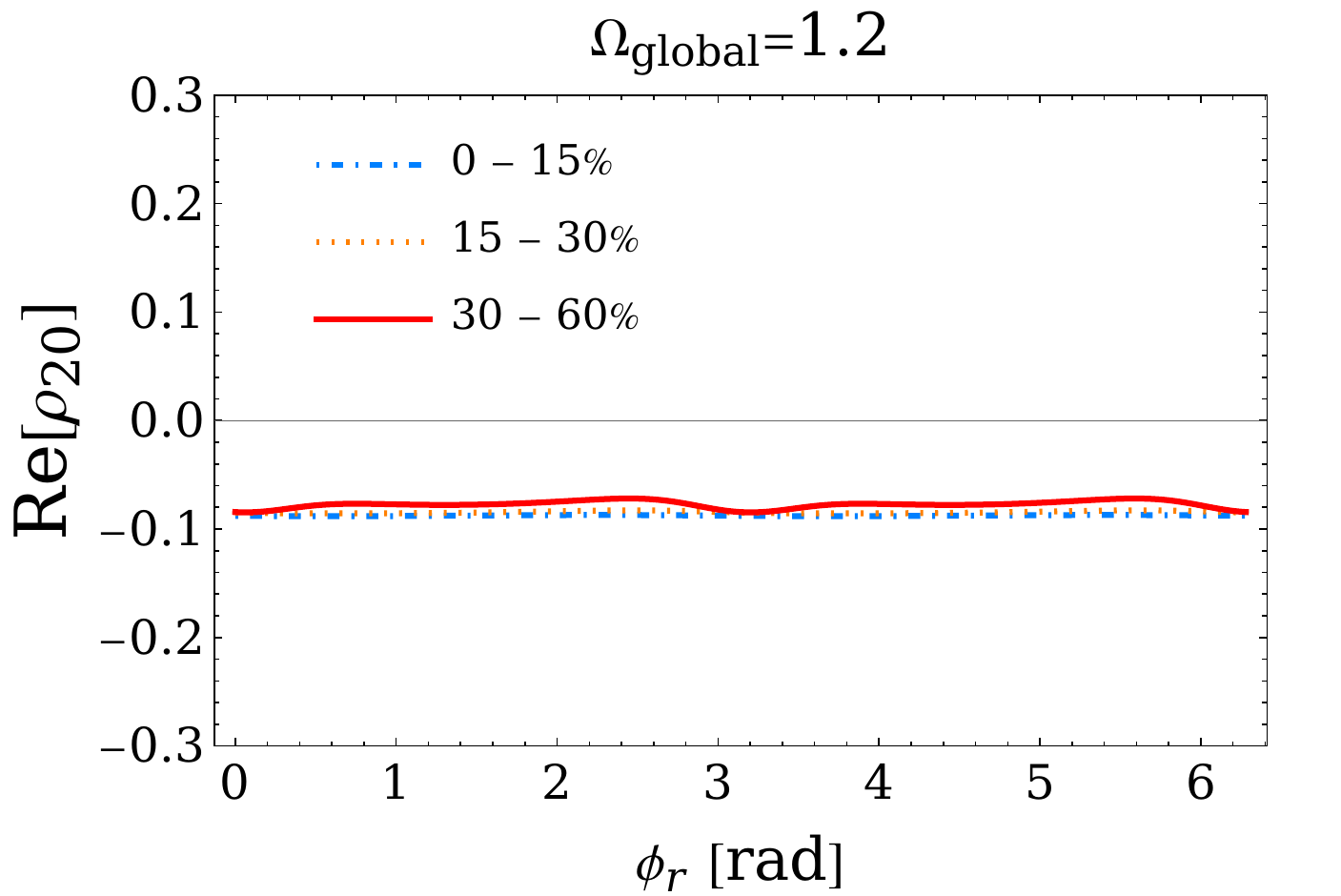}
   \epsfig{width=0.49\textwidth,figure=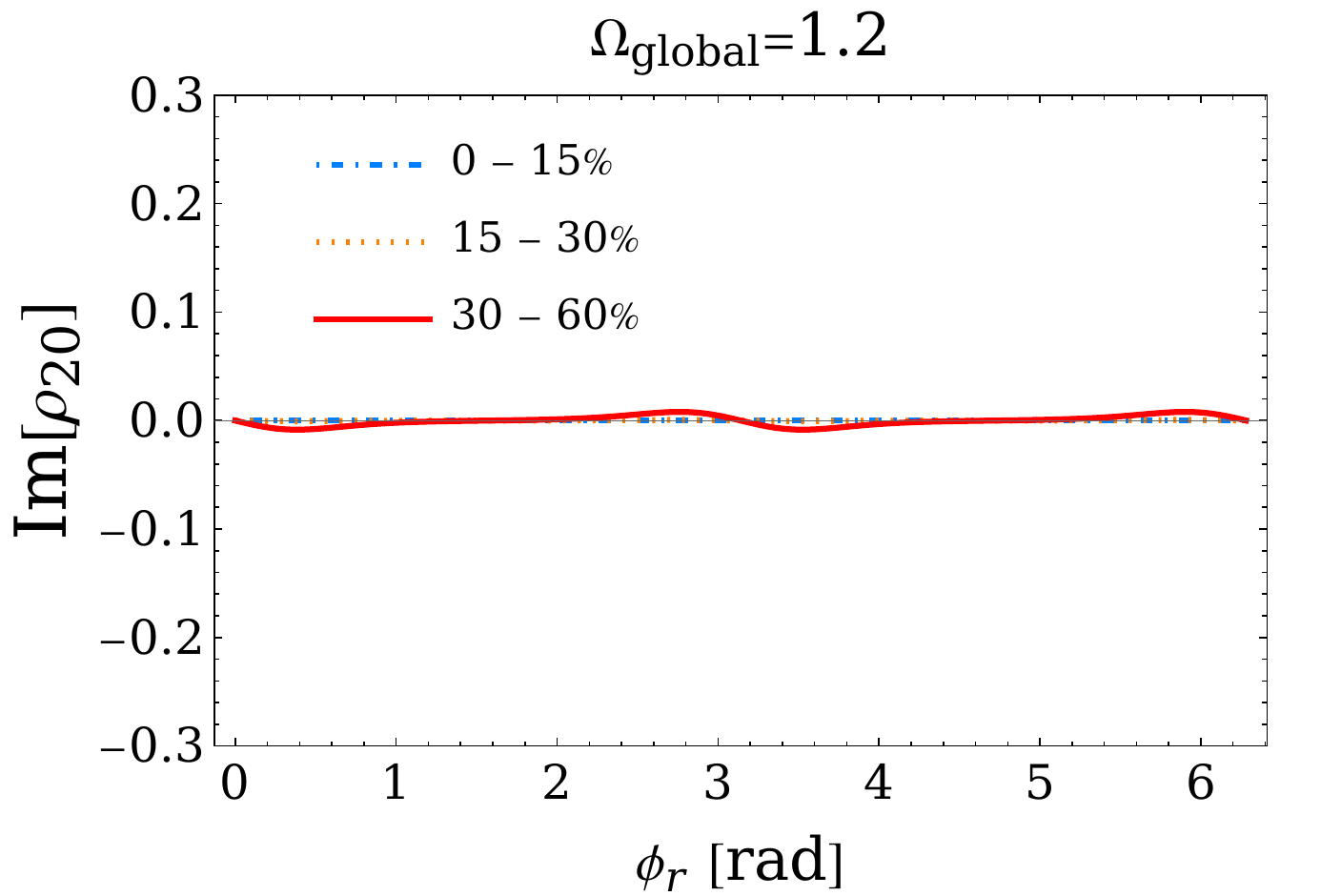} 
               \caption{ Non-zero $\rho_{|i|\ne |j|}$ matrix elements for zero and finite vorticity. $\rho_{0,-j}$ are the same as $\rho_{0,j}$ by symmetry. All other $\rho_{|i|\ne |j|}$ components of the density matrix, averaged over the space-time rapidity, were found to be zero. \label{offdiag}}
\end{figure*}

In this section, we present our results for the diagonal elements of the $f_2$ spin density matrix, as well as the real and imaginary parts of $\rho_{20}$, computed from Eq.~(\ref{density}) within the blast-wave model. The density matrix elements are expressed in terms of the spatial components of the thermal vorticity and the temperature; the explicit expressions for the alignment factors are provided in Appendix \ref{sec:f2dendiag}. We evaluated the mean values of diagonal elements by averaging over spacetime rapidity range $-4\leq\eta\leq4$ for three different centrality classes(0-15\%, 15-30\%, and 30-60\%). The results are shown in Figs.~\ref{fig:rho_coefficients_noglobal}, \ref{fig:rho_coefficients} and \ref{fig:rho_coefficients_largeomega}, which correspond to $\Omega_{\text{global}}=0$, $\Omega_{\text{global}}=0.024$ and $\Omega_{\text{global}}=1.2$, respectively. The mean values of real and imaginary part of $\rho_{20}$ were also obtained in the same manner and are shown in Fig.~\ref{offdiag}. The parameters used for our calculations are listed in Table \ref{table:parameters}. 

\begin{table}
\caption{Thermal model parameters used in our calculation \cite{florkblast,florkblast2}.}\label{table:parameters}
\begin{tabular}{c   c   c   c   c}
\hline\hline c\%&$ \epsilon$ & $\delta$ & $\tau_f$ (fm) & $r_{\text{max}}$ (fm)\\\hline 0-15 & 0.055 & 0.12 & 7.666 & 6.540\\15-30 & 0.097 & 0.26 & 6.258 & 5.417\\30-60 & 0.137 & 0.37 & 4.266 & 3.779\\\hline\hline
\end{tabular}
\end{table}
The oscillatory behavior of the density matrix element becomes distinct in more peripheral collisions, as they induce larger vorticity than head-on collisions. We also note that the addition of global vorticity perpendicular to the reaction plane increases the amplitude of $\rho_{00}$.

By symmetry, since in equilibrium there is only one angular momentum vector, observable combinations (\eqn{eq:f2dist}) involving $\rho_{\pm 2,\pm 1}$ and $\rho_{\pm 1,0}$ matrix elements are odd about this vector, and hence will be zero when integrated over the $\phi_r$ angle.
Thus, comparing \eqn{eq:f2dist} with Fig. \ref{offdiag} all observable combinations of the type $\rho_{|i|\neq |j|}$ elements are zero except $\rho_{2,0}$ (Shown in Fig \ref{offdiag}). This is however not true if spin and vorticity are not in equilibrium, i.e. scenarios such as \cite{montediss,montediss2,montediss3}, in which case 
the axial symmetry is broken and all density matrix elements are potentially non-zero.
$\rho_{|i|\ne |j|}$ density matrix elements are therefore good candidates for signatures for spin-vorticity at non-equilibrium and a dynamics such as \eqn{fertvalet}.  Estimates of the mixing, $1-\Tr\rho^2$ might also be used to rule out the statistical model considered here since it predicts a maximally mixed state.  However, since we do not have access to all density matrix elements, the mixing will at best be a broad confidence region.

While in this work we focused on the spin-0 decay of $f_2$ mesons, the recently measured $\chi_c$ spin-2 state decay into fermions \cite{cmschi} could also be used for a similar analysis, although of course the angular dependence and its relation to vorticity is expected to be different.

In conclusion, in this work we have shown that the density matrix structure of the $f_2$ meson decay into scalars is both rich in physics and experimentally accessible, which makes it a very promising probe of vorticity and spin dynamics in heavy ion collisions. We look forward to further experimental and phenomenological developments in this direction.

%\section{Summary}\label{sec:summary}In this work, we computed the angular distribution of produced pions in $f_2\to\pi+\pi$ decay, which is expressed in terms of angle between initial quantization axis and final particle momentum, and spin density matrix element of $f_2$ in its rest frame. The diagonal elements of $f_2$ spin density matrix, which are given as a function of an azimuthal angle w.r.t. impact parameter, were evaluated for different centrality classes in thermalized environment. The oscillatory amplitude increases as centrality increases. The addition of global vorticity along the axis perpendicular to the reaction plane increases the probability of finding $f_2$ in.

%\begin{figure*}[t]
%		\centering
    %\epsfig{width=0.49\textwidth,figure=fig_omega0/fig_r1_omega0.pdf}
   %\epsfig{width=0.49\textwidth,figure=fig_omega0/fig_alpha1_omega0.pdf}
   %\epsfig{width=0.49\textwidth,figure=fig_omega0/fig_r2_omega0.pdf}
   %\epsfig{width=0.49\textwidth,figure=fig_omega0/fig_alpha2_omega0.pdf}
   %\epsfig{width=0.49\textwidth,figure=fig_omega0/fig_r3_omega0.pdf}
   %\epsfig{width=0.49\textwidth,figure=fig_omega0/fig_alpha3_omega0.pdf}
    %        \caption{ .}
%\end{figure*}

\begin{acknowledgments}
The work of I. Park, G. Torrieri and S.H. Lee were supported by the Korean-Brazilian collaboration fund by the  Korea National Research
Foundation (NRF) under Grant No. 
RS-2023-NR121103. B. Kim and S. Lim acknowledge support from the National Research Foundation of Korea (NRF) grant funded by the Korean government (MSIT) under Contract No. NRF-2008-00458.  

\end{acknowledgments}

\clearpage
\appendix

\begin{widetext}

\section{\label{sec:twomethod}Calculation of general angular distribution of $f_2(1270)\to\pi+\pi$}

\subsection{\label{sec:spin2tensor}Interaction Lagrangian}

The polarization tensor of a massive spin-2 particle is obtained through a direct product between spin-1 polarization vectors. The Clebsch-Gordan(CG) coefficient has been taken from \cite{ParticleDataGroup:2024cfk}.

\begin{align}
\varepsilon^{\mu\nu}(\boldsymbol{0},\pm2)&=\varepsilon^{\mu}(\boldsymbol{0},\pm1)\varepsilon^{\nu}(\boldsymbol{0},\pm1)-\frac{1}{2}\begin{pmatrix}0&0&0&0\\0&1&\pm i&0\\0&\pm i&-1&0\\0&0&0&0\end{pmatrix},\nonumber\\
\varepsilon^{\mu\nu}(\boldsymbol{0},\pm1)&=\frac{1}{\sqrt{2}}\left(\varepsilon^{\mu}(\boldsymbol{0},\pm1)\varepsilon^{\nu}(\boldsymbol{0},0)+\varepsilon^{\mu}(\boldsymbol{0},0)\varepsilon^{\nu}(\boldsymbol{0},\pm1)\right)=\frac{1}{2}\begin{pmatrix}0&0&0&0\\0&0&0&\mp1\\0&0&0&-i\\0&\mp1&-i&0\end{pmatrix},\\
\varepsilon^{\mu\nu}(\boldsymbol{0},0)&=\frac{1}{\sqrt{6}}\left(\varepsilon^{\mu}(\boldsymbol{0},+1)\varepsilon^{\nu}(\boldsymbol{0},-1)+\varepsilon^{\mu}(\boldsymbol{0},-1)\varepsilon^{\nu}(\boldsymbol{0},+1)\right)+\sqrt{\frac{2}{3}}\varepsilon^{\mu}(\boldsymbol{0},0)\varepsilon^{\nu}(\boldsymbol{0},0)=\frac{1}{\sqrt{6}}\begin{pmatrix}0&0&0&0\\0&-1&0&0\\0&0&-1&0\\0&0&0&2\end{pmatrix}.\nonumber
\end{align}
The completeness relationship of spin-2 polarization tensor is given in \cite{Kim:2024mqx,Choi:1992ba}.

\begin{align}
\sum_{\lambda=-2}^{2}\varepsilon^{\mu\alpha}(\boldsymbol{p},\lambda)\varepsilon^{\nu\beta*}(\boldsymbol{p},\lambda)&=\tensor{\Delta}{^{\mu}^{\nu}^{\alpha}^{\beta}}(2,p)=\frac{1}{2}(\tensor{I}{^{\mu}^{\nu}}\tensor{I}{^{\alpha}^{\beta}}+\tensor{I}{^{\mu}^{\beta}}\tensor{I}{^{\nu}^{\alpha}})-\frac{1}{3}\tensor{I}{^{\mu}^{\alpha}}\tensor{I}{^{\nu}^{\beta}},\nonumber\\
I^{\mu\nu}&=-g^{\mu\nu}+\frac{p^{\mu}p^{\nu}}{p^2},\\
\varepsilon^{\mu\nu}(\boldsymbol{p},\lambda)&=\varepsilon^{\nu\mu}(\boldsymbol{p},\lambda),\;\;p_{\mu}\varepsilon^{\mu\nu}(\boldsymbol{p},\lambda)=0.\nonumber
\end{align}
Assume that the final pion momentum in $f_2$ rest frame is parameterized by $(\abs{\boldsymbol{p}_1},\theta,\phi)$ where $\theta$ and $\phi$ are polar and azimuthal angle with respect to the initial quantization axis. $\abs{\boldsymbol{p}_1}$ is given in Eq.~(\ref{eq:decaywidth}).
\begin{align}
p_1^{\mu}&=(\sqrt{\abs{\boldsymbol{p}_1}^2+m_{\pi}^2},\;\;\abs{\boldsymbol{p}_1}\sin\theta\cos\phi,\;\;\abs{\boldsymbol{p}_1}\sin\theta\sin\phi,\;\;\abs{\boldsymbol{p}_1}\cos\theta).
\end{align}
The squared of matrix element in Eq. (\ref{eq:generalf2}) is calculated to be

\begin{align}
\abs{\mathcal{M}}^2&=\frac{128\pi g_{f_{2}\pi\pi}^2}{15m_{f_2}^2}\abs{\boldsymbol{p}_1}^4W(\theta,\phi,\rho_{ij}).
\end{align}
$W(\theta,\phi,\rho_{ij})$ is given in the next subsection.

\subsection{\label{sec:helformalism}Helicity formalism}

The rotation in 3D space can be parameterized by Euler angles, $(\alpha,\beta,\gamma)$. We make a rotation around a fixed $z$-axis by an angle of $\gamma$ followed by a rotation around a fixed $y$-axis by an angle of $\beta$ finally followed by a rotation around a fixed $z$-axis by an angle of $\alpha$. The Wigner D-matrix is defined as below,
\begin{align}
D^J(\alpha,\beta,\gamma)_{MM^{\prime}}&=\mel{JM}{U\left(R(\alpha,\beta,\gamma)\right)}{JM^{\prime}}=e^{-i(M\alpha+M^{\prime}\gamma)}d^J(\beta)_{MM^{\prime}}.
\end{align}
More information about helicity formalism can be found in \cite{Leader:2001nas,devanathan2005angular,Chung:1971ri}. The angular distribution given in Eq. (\ref{eq:angdistformula}) can be rewritten in terms of trigonometric function as below,

\begin{eqnarray}
W(\theta,\phi,\rho_{ij})&=&\sum_{\lambda_1\lambda_2}\sum_{MM^{\prime}}\frac{2J+1}{4\pi}d^J(\theta)_{M\lambda}d^J(\theta)_{M^{\prime}\lambda}\abs{F^{J}_{\lambda_1\lambda_2}}^2\nonumber\\
&&\times\left(\Re[\rho_{MM^{\prime}}]\cos(M-M^{\prime})\phi-\Im[\rho_{MM^{\prime}}]\sin(M-M^{\prime})\phi\right).
\end{eqnarray}
As a result, the distribution of final pion produced in $f_2\to\pi+\pi$ is obtained by inserting $J=2$ in Eq.~(\ref{eq:angdistformula}) or Eq.~(\ref{eq:j00decay}).
\begin{eqnarray}\label{eq:f2dist}
W(\theta,\phi,\rho_{ij})&=&\frac{5}{64\pi}\bigg(\rho_{00}(3\cos2\theta+1)^{2}+6(\rho_{11}+\rho_{-1-1})\sin^{2}2\theta+6(\rho_{22}+\rho_{-2-2})\sin^{4}\theta\nonumber\\
&&-4\sqrt{6}\sin2\theta(3\cos^{2}\theta-1)\left(\Re[\rho_{10}-\rho_{0-1}]\cos\phi-\Im[\rho_{10}-\rho_{0-1}]\sin\phi\right)\nonumber\\
&&-12\sin2\theta\sin^{2}\theta\left(\Re[\rho_{21}-\rho_{-1-2}]\cos\phi-\Im[\rho_{21}-\rho_{-1-2}]\sin\phi\right)\nonumber\\
&&+\sqrt{6}(3\sin^{2}2\theta-4\sin^{2}\theta)\left(\Re[\rho_{20}+\rho_{0-2}]\cos2\phi-\Im[\rho_{20}+\rho_{0-2}]\sin2\phi\right)\nonumber\\
&&-12\sin^{2}2\theta\left(\Re[\rho_{1-1}]\cos2\phi-\Im[\rho_{1-1}]\sin2\phi\right)\nonumber\\
&&+12\sin2\theta\sin^{2}\theta\left(\Re[\rho_{2-1}-\rho_{1-2}]\cos3\phi-\Im[\rho_{2-1}-\rho_{1-2}]\sin3\phi\right)\nonumber\\
&&+12\sin^{4}\theta\left(\Re[\rho_{2-2}]\cos4\phi-\Im[\rho_{2-2}]\sin4\phi\right)\bigg).
\end{eqnarray}
The integration over azimuthal angle $\phi$ of the angular distribution given in Eq.~(\ref{eq:j00decay}) yields the distribution in terms of polar angle $\theta$ and diagonal spin density matrix elements.
\begin{eqnarray}
W(\theta,\rho_{ii})&=&\frac{5}{12}\Big(\rho_{11}+\rho_{-1-1}+2(\rho_{22}+\rho_{-2-2})+3(\rho_{11}+\rho_{-1-1}-\rho_{22}-\rho_{-2-2})P_1(\cos\theta)^2\nonumber\\
&&+\left(\rho_{22}+\rho_{-2-2}-4(\rho_{11}+\rho_{-1-1})+6\rho_{00}\right)P_2(\cos\theta)^2\Big).
\end{eqnarray}
$P_J(\cos\theta)$ is Legendre polynomial. The formula in terms of $\cos\theta$ is given in Eq.~(\ref{wonlydiag}).

\section{\label{sec:f2dendiag}Diagonal density matrix components for the $f_2$ tensor meson}
Using the density matrix definition of the $f_2$ tensor meson in Eq.~\eqref{density}, and assuming local thermodynamic equilibrium, we can explicitly obtain all diagonal elements of its density matrix, where the Pauli matrix for $S=2$ are:
\begin{equation}
    S_x=\left(
\begin{array}{ccccc}
 0 & 1 & 0 & 0 & 0 \\
 1 & 0 & \sqrt{\frac{3}{2}} & 0 & 0 \\
 0 & \sqrt{\frac{3}{2}} & 0 & \sqrt{\frac{3}{2}}
   & 0 \\
 0 & 0 & \sqrt{\frac{3}{2}} & 0 & 1 \\
 0 & 0 & 0 & 1 & 0 \\
\end{array}
\right),\;\;\; S_y = \left(
\begin{array}{ccccc}
 0 & -i & 0 & 0 & 0 \\
 i & 0 & -i \sqrt{\frac{3}{2}} & 0 & 0 \\
 0 & i \sqrt{\frac{3}{2}} & 0 & -i
   \sqrt{\frac{3}{2}} & 0 \\
 0 & 0 & i \sqrt{\frac{3}{2}} & 0 & -i \\
 0 & 0 & 0 & i & 0 \\
\end{array}
\right),\;\;\;  S_z = \left(
\begin{array}{ccccc}
 2 & 0 & 0 & 0 & 0 \\
 0 & 1 & 0 & 0 & 0 \\
 0 & 0 & 0 & 0 & 0 \\
 0 & 0 & 0 & -1 & 0 \\
 0 & 0 & 0 & 0 & -2 \\
\end{array}
\right).
\label{spin2}
\end{equation}
From the definition of the spatial components of thermal vorticity Eq.(\ref{spacial_vorticity}) (it is important to note that $\frac{\vec{\Omega}}{T} \xrightarrow{}\vec{\Omega}$ in order to maintain consistency with the definition of thermal vorticity) together of density matrix Eq.~(\ref{density}) and Pauli matrix to $S=2$, Eq.~(\ref{spin2}) we can obtain the trace of density matrix:
\begin{equation}
    \Omega = \sqrt{\Omega _1^2+\Omega
   _2^2+\Omega _3^2},
\end{equation}

\begin{equation}
   \mathrm{Tr}(\rho)=Z= 1+2 \cosh \left(\Omega\right)+2 \cosh \left(2
   \Omega\right).
\end{equation}

\begin{description}
\item[The coefficient] $\rho_{\pm2\pm2}\left(\Omega_1,\Omega_2,\Omega_3, T\right)$:
\begin{equation}
\rho_{\pm2\pm2}\left(\Omega_1,\Omega_2,\Omega_3, T\right)= \frac{1}{Z\Omega^4}\left(\Omega _1^4 \cosh ^4\left(\frac{\Omega}{2}
   \right)+\Omega _2^4 \cosh
   ^4\left(\frac{\Omega}{2}\right)+\Omega _1^2 \cosh
   ^2\left(\frac{\Omega}{2}\right) \left(\Omega _2^2-2
   \Omega _3^2\pm2 \Omega _3 \Omega \sinh
   \left(\Omega\right)+\right.\right.
\end{equation}
\[
    \left.\left. +\left(\Omega _2^2+4 \Omega
   _3^2\right) \cosh \left(\Omega\right)\right)+2
   \Omega _2^2 \Omega _3 \cosh
   ^2\left(\frac{\Omega}{2}\right) \left(\Omega _3 \left(2 \cosh
   \left(\Omega\right)-1\right)\pm\Omega \sinh
   \left(\Omega\right)\right)+\Omega _3^3
   \left(\Omega _3 \cosh
   \left(2 \Omega\right)\pm\Omega \sinh \left(2 \Omega\right)\right)\right).
\]
\item[The coefficient] $\rho_{\pm1\pm1}\left(\Omega_1,\Omega_2,\Omega_3, T\right)$:
\begin{equation}
   \rho_{\pm1\pm1}\left(\Omega_1,\Omega_2,\Omega_3,T\right)= \frac{1}{2Z\Omega^4}\left(\left(\Omega _3^2+\left(\Omega _1^2+\Omega
   _2^2\right) \left(2 \cosh \left(\Omega\right)-1\right)\right) \left(\Omega
   _1^2+\Omega _2^2\pm2 \Omega _3 \Omega \sinh
   \left(\Omega\right)+\right.\right.
\end{equation}
\[
    \left.\left.+\left(\Omega _1^2+\Omega _2^2+2
   \Omega _3^2\right) \cosh \left(\Omega\right)\right)\right).
\]
\item[The coefficient] $\rho_{00}\left(\Omega_1,\Omega_2,\Omega_3, T\right)$:

\begin{equation}
    \rho_{00}\left(\Omega_1,\Omega_2,\Omega_3, T\right)= \frac{1}{4Z\Omega^4}\left(3 \left(\Omega _1^2+\Omega
   _2^2\right){}^2\cosh(2\Omega)+12 \Omega _3^2
   \left(\Omega _1^2+\Omega _2^2\right)\cosh(\Omega)+\left(\Omega _1^2+\Omega_2^2-2 \Omega _3^2\right)^2\right).
\end{equation}

\item[The coefficient] $\Re[\rho_{10}-\rho_{0-1}]$:

\begin{equation}
    \Re[\rho_{10}-\rho_{0-1}]= \frac{2 \sqrt{6} \Omega _1 \Omega _3 \sinh
   ^2\left(\frac{\Omega}{2}\right) \left(\Omega
   _3^2+\left(\Omega _1^2+\Omega _2^2\right)
   \cosh \left(\Omega\right)\right)}{Z\Omega^4}
\end{equation}

\item[The coefficient] $\Im[\rho_{10}-\rho_{0-1}]$:
\begin{equation}
    \Im[\rho_{10}-\rho_{0-1}]= -\frac{2 \sqrt{6} \Omega _2 \Omega _3 \sinh
   ^2\left(\frac{\Omega}{2}\right) \left(\Omega
   _3^2+\left(\Omega _1^2+\Omega _2^2\right)
   \cosh \left(\Omega\right)\right)}{Z\Omega^4}.
\end{equation}
\item[The coefficient] $\Re[\rho_{21}-\rho_{-1-2}]$:
\begin{equation}
    \Re[\rho_{21}-\rho_{-1-2}]= \frac{2 \Omega _1 \Omega _3 \sinh
   ^2\left(\frac{\Omega}{2} \right) \left(3 \Omega
   _1^2+3 \Omega _2^2+2 \Omega _3^2+\left(4
   \Omega _3^2+3 \left(\Omega _1^2+\Omega
   _2^2\right)\right) \cosh \left(\Omega\right)\right)}{Z\Omega^4}.
\end{equation}

\item[The coefficient] $\Im[\rho_{21}-\rho_{-1-2}]$:
\begin{equation}
    \Im[\rho_{21}-\rho_{-1-2}]= \frac{2 \Omega _2 \Omega _3 \sinh
   ^2\left(\frac{\Omega}{2} \right) \left(-3 \Omega
   _1^2-3 \Omega _2^2-2 \Omega _3^2-\left(4
   \Omega _3^2+3 \left(\Omega _1^2+\Omega
   _2^2\right)\right) \cosh \left(\Omega\right)\right)}{Z\Omega^4}.
\end{equation}

\item[The coefficient] $\Re[\rho_{20}+\rho_{0-2}]$:
\begin{equation}
    \Re[\rho_{20}+\rho_{0-2}]= \frac{\sqrt{6} \left(\Omega _1^2-\Omega
   _2^2\right) \sinh ^2\left(\frac{\Omega}{2}
   \right) \left(\Omega _1^2+\Omega
   _2^2+\left(\Omega _1^2+\Omega _2^2+2 \Omega
   _3^2\right) \cosh \left(\Omega\right)\right)}{Z\Omega^4}.
\end{equation}

\item[The coefficient] $\Im[\rho_{20}+\rho_{0-2}]$:
\begin{equation}
    \Im[\rho_{20}+\rho_{0-2}]= -\frac{2 \sqrt{6} \Omega _1 \Omega _2 \sinh
   ^2\left(\frac{\Omega}{2}\right) \left(\Omega
   _1^2+\Omega _2^2+\left(\Omega _1^2+\Omega
   _2^2+2 \Omega _3^2\right) \cosh
   \left(\Omega\right)\right)}{Z\Omega^4}.
\end{equation}

\item[The coefficient] $\Re[\rho_{1-1}]$:
\begin{equation}
    \Re[\rho_{1-1}]= \frac{\left(\Omega _1^2-\Omega _2^2\right) \sinh
   ^2\left(\frac{\Omega}{2}\right) \left(3 \Omega
   _3^2+\left(\Omega _1^2+\Omega _2^2\right)
   \left(2 \cosh \left(\Omega\right)+1\right)\right)}{Z\Omega^4}.
\end{equation}

\item[The coefficient] $\Im[\rho_{1-1}]$:
\begin{equation}
    \Im[\rho_{1-1}]= -\frac{2 \Omega _1 \Omega _2 \sinh
   ^2\left(\frac{\Omega}{2} \right) \left(3 \Omega
   _3^2+\left(\Omega _1^2+\Omega _2^2\right)
   \left(2 \cosh \left(\Omega\right)+1\right)\right)}{Z\Omega^4}.
\end{equation}

\item[The coefficient] $\Re[\rho_{2-1}-\rho_{1-2}]$:
\begin{equation}
   \Re[\rho_{2-1}-\rho_{1-2}]=\frac{4 \Omega _1 \left(\Omega _1^2-3 \Omega
   _2^2\right) \Omega _3 \sinh
   ^4\left(\frac{\Omega}{2}\right)}{Z\Omega^4} .
\end{equation}

\item[The coefficient] $\Im[\rho_{2-1}-\rho_{1-2}]$:
\begin{equation}
    \Im[\rho_{2-1}-\rho_{1-2}]= \frac{4 \Omega _2 \left(\Omega _2^2-3 \Omega
   _1^2\right) \Omega _3 \sinh
   ^4\left(\frac{\Omega}{2}\right)}{Z\Omega^4}.
\end{equation}

\item[The coefficient] $\Re[\rho_{2-2}]$:
\begin{equation}
    \Re[\rho_{2-2}]= \frac{\left(\Omega _1^4-6 \Omega _2^2 \Omega
   _1^2+\Omega _2^4\right) \sinh
   ^4\left(\frac{\Omega}{2} \right)}{Z\Omega^4}.
\end{equation}

\item[The coefficient] $\Im[\rho_{2-2}]$:
\begin{equation}
    \Im[\rho_{2-2}]= \frac{4 \Omega _1 \Omega _2 \left(\Omega
   _2^2-\Omega _1^2\right) \sinh
   ^4\left(\frac{\Omega}{2}\right)}{Z\Omega^4}.
\end{equation}

We can obtain the mean value of diagonal elements over the space-time rapidity in following way:
\begin{equation}
    \left<\rho_{i,j}(\phi_r)\right> = \frac{1}{\eta-\eta_0}\int_{\eta_0}^{\eta} d\eta'\rho_{i,j}(\phi_r,\eta'),
\end{equation}
\end{description}
where $\eta=4$ and $\eta_0=-4$.

\end{widetext}

\bibliography{reference}% Produces the bibliography via BibTeX.

@article{STAR:2017ckg,
    author = "Adamczyk, L. and others",
    collaboration = "STAR",
    title = "{Global $\Lambda$ hyperon polarization in nuclear collisions: evidence for the most vortical fluid}",
    eprint = "1701.06657",
    archivePrefix = "arXiv",
    primaryClass = "nucl-ex",
    doi = "10.1038/nature23004",
    journal = "Nature",
    volume = "548",
    pages = "62--65",
    year = "2017"
}

@article{ALICE:2025cdf,
    author = "Acharya, Shreyasi and others",
    collaboration = "ALICE",
    title = "{First measurement of D$^{*+}$ vector meson spin alignment in Pb{\textendash}Pb collisions at $\sqrt{{s}_{\text{NN}}}={5}.0{2}$ TeV}",
    eprint = "2504.00714",
    archivePrefix = "arXiv",
    primaryClass = "nucl-ex",
    reportNumber = "CERN-EP-2025-072",
    doi = "10.1007/JHEP10(2025)094",
    journal = "JHEP",
    volume = "10",
    pages = "094",
    year = "2025"
}

@report{cmschi,
    collaboration = "CMS",
    title = "{Study of $\chi_\mathrm{c}$ production in pPb collisions at $\sqrt{\smash[b]{s_{\mathrm{NN}}}}=8.16\,\text{TeV}$ with the CMS experiment}",
    reportNumber = "CMS-PAS-HIN-22-003",
    year = "2025"
}

@article{wangspin,
    author = "Liang, Zuo-Tang and Wang, Xin-Nian",
    title = "{Spin alignment of vector mesons in non-central A+A collisions}",
    eprint = "nucl-th/0411101",
    archivePrefix = "arXiv",
    reportNumber = "LBNL-56659",
    doi = "10.1016/j.physletb.2005.09.060",
    journal = "Phys. Lett. B",
    volume = "629",
    pages = "20--26",
    year = "2005"
}

@article{ourpaper,
    author = "Montenegro, David and Torrieri, Giorgio",
    title = "{Causality and dissipation in relativistic polarizable fluids}",
    eprint = "1807.02796",
    archivePrefix = "arXiv",
    primaryClass = "hep-th",
    doi = "10.1103/PhysRevD.100.056011",
    journal = "Phys. Rev. D",
    volume = "100",
    number = "5",
    pages = "056011",
    year = "2019"
}

@article{ALICE:2019aid,
    author = "Acharya, Shreyasi and others",
    collaboration = "ALICE",
    title = "{Evidence of Spin-Orbital Angular Momentum Interactions in Relativistic Heavy-Ion Collisions}",
    eprint = "1910.14408",
    archivePrefix = "arXiv",
    primaryClass = "nucl-ex",
    reportNumber = "CERN-EP-2019-251",
    doi = "10.1103/PhysRevLett.125.012301",
    journal = "Phys. Rev. Lett.",
    volume = "125",
    number = "1",
    pages = "012301",
    year = "2020"
}

@article{Khatun:2024vgn,
    author = "Khatun, Anisa",
    collaboration = "ALICE",
    title = "{UPC physics with ALICE in Run 3}",
    eprint = "2405.19069",
    archivePrefix = "arXiv",
    primaryClass = "hep-ex",
    doi = "10.17161/5b46x852",
    journal = "Phys. Proc. UPC",
    volume = "1",
    pages = "22",
    year = "2024"
}

@article{kaminski,
    author = "Hongo, Masaru and Huang, Xu-Guang and Kaminski, Matthias and Stephanov, Mikhail and Yee, Ho-Ung",
    title = "{Relativistic spin hydrodynamics with torsion and linear response theory for spin relaxation}",
    eprint = "2107.14231",
    archivePrefix = "arXiv",
    primaryClass = "hep-th",
    reportNumber = "RIKEN-iTHEMS-Report-21",
    doi = "10.1007/JHEP11(2021)150",
    journal = "JHEP",
    volume = "11",
    pages = "150",
    year = "2021"
}

@article{montediss,
    author = "Montenegro, David and Torrieri, Giorgio",
    title = "{Linear response theory and effective action of relativistic hydrodynamics with spin}",
    eprint = "2004.10195",
    archivePrefix = "arXiv",
    primaryClass = "hep-th",
    doi = "10.1103/PhysRevD.102.036007",
    journal = "Phys. Rev. D",
    volume = "102",
    number = "3",
    pages = "036007",
    year = "2020"
}

@article{montediss2,
    author = "Torrieri, Giorgio and Montenegro, David",
    title = "{Linear response hydrodynamics of a relativistic dissipative fluid with spin}",
    eprint = "2207.00537",
    archivePrefix = "arXiv",
    primaryClass = "hep-th",
    doi = "10.1103/PhysRevD.107.076010",
    journal = "Phys. Rev. D",
    volume = "107",
    number = "7",
    pages = "076010",
    year = "2023"
}

@article{montediss3,
    author = "Montenegro, David and Torrieri, Giorgio",
    title = "{Causality of polarizable dissipative fluids from Lagrangian hydrodynamics}",
    eprint = "2506.17327",
    archivePrefix = "arXiv",
    primaryClass = "nucl-th",
    doi = "10.1103/yjtd-qztf",
    journal = "Phys. Rev. D",
    volume = "112",
    number = "3",
    pages = "034041",
    year = "2025"
}

@article{Xia:2020tyd,
    author = "Xia, Xiao-Liang and Li, Hui and Huang, Xu-Guang and Zhong Huang, Huan",
    title = "{Local spin alignment of vector mesons in relativistic heavy-ion collisions}",
    eprint = "2010.01474",
    archivePrefix = "arXiv",
    primaryClass = "nucl-th",
    doi = "10.1016/j.physletb.2021.136325",
    journal = "Phys. Lett. B",
    volume = "817",
    pages = "136325",
    year = "2021"
}

@article{kayman1,
    author = "Gon{\c{c}}alves, Kayman J. and Torrieri, Giorgio",
    title = "{Spin alignment of vector mesons as a probe of spin hydrodynamics and freeze-out}",
    eprint = "2104.12941",
    archivePrefix = "arXiv",
    primaryClass = "nucl-th",
    doi = "10.1103/PhysRevC.105.034913",
    journal = "Phys. Rev. C",
    volume = "105",
    number = "3",
    pages = "034913",
    year = "2022"
}

@article{kayman2,
    author = "De Moura, Paulo Henrique and Goncalves, Kayman J. and Torrieri, Giorgio",
    title = "{Quarkonium spin alignment in a vortical medium}",
    eprint = "2305.02985",
    archivePrefix = "arXiv",
    primaryClass = "hep-ph",
    doi = "10.1103/PhysRevD.108.034032",
    journal = "Phys. Rev. D",
    volume = "108",
    number = "3",
    pages = "034032",
    year = "2023"
}

@article{STAR:2022fan,
    author = "Abdallah, M. S. and others",
    collaboration = "STAR",
    title = "{Pattern of global spin alignment of {\ensuremath{\phi}} and K$^{*0}$ mesons in heavy-ion collisions}",
    eprint = "2204.02302",
    archivePrefix = "arXiv",
    primaryClass = "hep-ph",
    doi = "10.1038/s41586-022-05557-5",
    journal = "Nature",
    volume = "614",
    number = "7947",
    pages = "244--248",
    year = "2023"
}

@article{Shklyar:2004ba,
    author = "Shklyar, V. and Lenske, H. and Mosel, U. and Penner, G.",
    title = "{Coupled-channel analysis of the omega-meson production in pi N and gamma N reactions for c.m. energies up to 2-GeV}",
    eprint = "nucl-th/0412029",
    archivePrefix = "arXiv",
    doi = "10.1103/PhysRevC.72.019903",
    journal = "Phys. Rev. C",
    volume = "71",
    pages = "055206",
    year = "2005",
    note = "[Erratum: Phys.Rev.C 72, 019903 (2005)]"
}

@article{Suzuki:1993zs,
    author = "Suzuki, M.",
    title = "{Tensor meson dominance: Phenomenology of the f2 meson}",
    doi = "10.1103/PhysRevD.47.1043",
    journal = "Phys. Rev. D",
    volume = "47",
    pages = "1043--1047",
    year = "1993"
}

@article{ParticleDataGroup:2024cfk,
    author = "Navas, S. and others",
    collaboration = "Particle Data Group",
    title = "{Review of particle physics}",
    doi = "10.1103/PhysRevD.110.030001",
    journal = "Phys. Rev. D",
    volume = "110",
    number = "3",
    pages = "030001",
    year = "2024"
}

@article{florkblast,
    author = "Florkowski, Wojciech and Kumar, Avdhesh and Ryblewski, Radoslaw and Mazeliauskas, Aleksas",
    title = "{Longitudinal spin polarization in a thermal model}",
    eprint = "1904.00002",
    archivePrefix = "arXiv",
    primaryClass = "nucl-th",
    doi = "10.1103/PhysRevC.100.054907",
    journal = "Phys. Rev. C",
    volume = "100",
    number = "5",
    pages = "054907",
    year = "2019"
}

@article{florkblast2,
    author = "Florkowski, Wojciech and Kumar, Avdhesh and Mazeliauskas, Aleksas and Ryblewski, Radoslaw",
    title = "{Effect of thermal shear on longitudinal spin polarization in a thermal model}",
    eprint = "2112.02799",
    archivePrefix = "arXiv",
    primaryClass = "hep-ph",
    doi = "10.1103/PhysRevC.105.064901",
    journal = "Phys. Rev. C",
    volume = "105",
    number = "6",
    pages = "064901",
    year = "2022"
}

@article{karpenko,
    author = "Becattini, F. and Karpenko, I. and Lisa, M. and Upsal, I. and Voloshin, S.",
    title = "{Global hyperon polarization at local thermodynamic equilibrium with vorticity, magnetic field and feed-down}",
    eprint = "1610.02506",
    archivePrefix = "arXiv",
    primaryClass = "nucl-th",
    doi = "10.1103/PhysRevC.95.054902",
    journal = "Phys. Rev. C",
    volume = "95",
    number = "5",
    pages = "054902",
    year = "2017"
}

@article{ryb,
    author = "Gon{\c{c}}alves, Kayman J. and Torrieri, Giorgio and Ryblewski, Radoslaw",
    title = "{Meson spin alignment and baryon polarization from coalescence with spin-vorticity nonequilibrium}",
    eprint = "2410.16448",
    archivePrefix = "arXiv",
    primaryClass = "hep-ph",
    doi = "10.1103/gsv6-zydk",
    journal = "Phys. Rev. C",
    volume = "112",
    number = "1",
    pages = "014901",
    year = "2025"
}

@article{Kim:2024mqx,
    author = "Kim, Sang-Ho and Oh, Yongseok and Son, Sangyeong and Sakinah, S. and Cheoun, Myung-Ki",
    title = "{Effective Lagrangian for strong and electromagnetic interactions of high-spin resonances}",
    eprint = "2412.18927",
    archivePrefix = "arXiv",
    primaryClass = "hep-ph",
    doi = "10.1103/PhysRevD.111.054031",
    journal = "Phys. Rev. D",
    volume = "111",
    number = "5",
    pages = "054031",
    year = "2025"
}

@article{Choi:1992ba,
    author = "Choi, S. Y. and Lee, Jungil and Shim, J. S. and Song, H. S.",
    title = "{Spin-2 particle polarization}",
    reportNumber = "SNUTP-92-27",
    journal = "J. Korean Phys. Soc.",
    volume = "25",
    pages = "576--579",
    year = "1992"
}

@book{Leader:2001nas,
    author = "Leader, Elliot",
    title = "{Spin in Particle Physics}",
    doi = "10.1017/cbo9780511524455",
    isbn = "978-1-009-40204-0, 978-1-009-40199-9, 978-1-009-40201-9, 978-0-511-87418-5, 978-0-521-35281-9, 978-0-521-02077-0",
    publisher = "Cambridge University Press",
    volume = "15",
    month = "7",
    year = "2001"
}

@book{devanathan2005angular,
  title={Angular Momentum Techniques in Quantum Mechanics},
  author={Devanathan, V.},
  isbn={9780306471230},
  lccn={99038094},
  series={Fundamental Theories of Physics},
  url={https://books.google.co.kr/books?id=KYThBwAAQBAJ},
  year={2005},
  publisher={Springer Netherlands}
}

@techreport{Chung:1971ri,
    author = "Chung, Suh Urk",
    title = "{SPIN FORMALISMS}",
    institution = "CERN",
    number = "CERN-71-08",
    doi = "10.5170/CERN-1971-008",
    month = "3",
    year = "1971"
}

\end{document}